\documentclass[journal]{IEEEtran}
\usepackage{graphicx}
\usepackage{subcaption} 
\usepackage{amssymb}
\usepackage{amsmath}
\usepackage{nccmath}
\usepackage{amsfonts}
\usepackage{float}
\usepackage{multirow}
\usepackage{array}
\usepackage{mfirstuc}
\MFUnocap{are}
\MFUnocap{or}
\MFUnocap{etc}
\usepackage{array}
\usepackage{xcolor}
\newcolumntype{L}{>{\centering\arraybackslash}m{1.5cm}}
\usepackage{cite}

\ifCLASSINFOpdf
\else
\fi

\begin{document}
%
\title{Exploring DMD-type Algorithms for Modeling Signalised Intersections}
%
%
%

\author{Kazi Redwan Shabab*,
        Shakib Mustavee*,
        Shaurya Agarwal,
        Mohamed H. Zaki,
        and Sajal Das

%
%

\thanks{* First and second authors contributed equally.}
\thanks{K.R. Shabab, S. Mustavee, S. Agarwal, and  M. Zaki are with University of Central Florida} 
\thanks{S. Das is with Missouri University of Science and Technology} 
\thanks{Manuscript received Nov 2020}}

%



\maketitle

\begin{abstract}
This paper explores a novel data-driven approach based on recent developments in Koopman operator theory and dynamic mode decomposition (DMD) for modeling signalized intersections. Vehicular flow and queue formation on signalized intersections have complex nonlinear dynamics, making system identification, modeling, and controller design tasks challenging. We employ a Koopman theoretic approach to transform the original nonlinear dynamics into locally linear infinite-dimensional dynamics. The data-driven approach relies entirely on spatio-temporal snapshots of the traffic data. We investigate several key aspects of the approach and provide insights into the usage of DMD-type algorithms for application in adaptive signalized intersections. To demonstrate the utility of the obtained linearized dynamics, we perform prediction of the queue lengths at the intersection; and compare the results with the state-of-the-art long short term memory (LSTM) method. The case study involves the morning peak vehicle movements and queue lengths at two Orlando area signalized intersections.  It is observed that DMD-based algorithms are able to capture complex dynamics with a linear approximation to a reasonable extent. 
\end{abstract}

\begin{IEEEkeywords}
Modeling, System Identification, Dynamic mode decomposition, Koopman operator theory, Adaptive signalized intersection, Queue length prediction.
\end{IEEEkeywords}



\IEEEpeerreviewmaketitle

\section{Introduction}
 \IEEEPARstart{D}{ata-driven} intelligent transportation systems (ITS) are increasingly playing a critical role in improving the efficiency of the existing transportation network and addressing, along the way, many challenges of the traffic in large cities such as safety and road congestion. Roadside and in-vehicle sensors are, in particular, a good example of the successful deployment of new technologies in transportation.  Automobile manufacturers are equipping new vehicles with various sensors for traffic awareness, safety, and communication purposes. Simultaneously, state agencies have started deploying infrastructure sensors such as cameras, detectors, and microwave radars coupled with communication devices such as roadside units (RSUs) \cite{haoui2008wireless,ma2017spatial}. These new-age traffic sensors provide an unprecedented amount of big data ready to be mined and utilized for data-driven ITS applications \cite{khattak1994concept,canepa2017networked,jin2018hybrid,ran2012perspectives}. In comparison to conventional data, big data is more capable of revealing the underlying traffic dynamics \cite{yaqoob2016big,zhao2019geographical, ran2012perspectives}. Big data has been used for traffic congestion detection \cite{cardenas2016traffic}, travel pattern identification \cite{liu2018big}, traffic flow prediction \cite{chen2018traffic,contreras2015observability, agarwal2015dynamic,ban2010performance,wang2005empirical}, crash data analysis \cite{arvin2020safety,yuan2018real,agarwal2016hybrid}, adaptive signal control \cite{muresan2019adaptive,mannion2016experimental,szeto2005impact}, ramp control  \cite{agarwal2015feedback}, and other related transportation research \cite{rahman2021real}.\\ 
The big data-based statistical models and machine learning algorithms are efficient in predicting a dynamical system. However, these are `black box' approaches that cannot capture or utilize the system's underlying dynamics. On the other hand, physics-based modeling often falls short of capturing uncertainty and complexity in a meaningful model that can be used to design controllers. The observed traffic dynamics are highly nonlinear and stochastic, making it challenging to obtain valuable models for ramp metering, signalized intersections and even more challenging to integrate them with existing dynamic traffic assignment (DTA) models. Uncovering complex traffic dynamics from high-fidelity data requires a paradigm shift in modeling dynamic transportation networks and incorporating them coherently with emerging technologies and existing models.\\
DMD theory has close connections with the Koopman operator theory. Koopman operator is an infinite-dimensional linear operator capable of describing the time evolution of system observables (measurements). The spectral properties of Koopman operator aid in identifying complex system dynamics' inherent properties by decomposition it into spatio-temporal coherent structures. The Koopman operator's triple decomposition into the eigenvalues, eigenfunctions, and Koopman modes is referred to as Koopman mode decomposition (KMD).
There are two main classes of computing KMD from real-world system measurements numerically --- Dynamic Mode Decomposition (DMD)  and Arnoldi Method.\\
Dynamic mode decomposition (DMD) is a purely data-driven technique that can provide a locally linear description of complex nonlinear dynamics. The noteworthy point about the method is that it does not require any prior information of the system or its internal physics to capture its dynamics, making it comparable to grey box models of system identification. Unlike purely data-driven statistical models and machine learning algorithms, DMD and related algorithms provide an approximate system identification. The identification of complex dynamical systems as approximate linear dynamics has several benefits. Among them is the simplicity in understanding the system and the applicability of linear control algorithms.\\
This paper explores two variants of dynamic mode decomposition--DMD with control (DMDc) and Hankel DMD with control (HDMDc)--for system identification and localized linear approximation of the dynamics. To demonstrate the application and validity of the obtained linearized dynamics, we predict the queue lengths at the intersection and compare the results with the state-of-the-art long short-term memory (LSTM) method. We also perform a case study involving the queue lengths at two signalized intersections in the Orlando metropolitan area. \vspace{1mm}\\
\noindent\textit{Contributions:} Proper system identification of queue length at signalized-intersection can play a pivotal role in  designing  optimal signal timing controller. This paper explores recent developments in the DMD algorithm and the associated Koopman theory for application in signalized intersections. There have been a few studies attempting to utilize DMD-based algorithms for applications in ITS \cite{ling2018koopman,ling2018operator, ling2020koopman}. However, this study adopts a different approach, as demonstrated by the following contributions. We propose a system identification of nonlinear traffic dynamics at signalized intersections using DMDc and HDMDc algorithms. We perform the queue length prediction as a surrogate measure to validate the system identification results - and compare them with the state-of-the-art LSTM approach. We also comment on the issue of an optimal number of training snapshots, the impact of delay embedding, and prediction window.  \vspace{1mm}


\textit{Outline:} The rest of the paper is organized as follows: Section-II discusses the existing literature on DMD, DMDc, and Hankel DMD, along with their applications. Section-III provides the required mathematical preliminaries of DMD, DMDc, and HDMDc theory. Section-IV describes the field data used in this study. Section-V provides details on the simulation model development and calibration. In section-VI, we formulate the problem and provide details of the DMDc application. Section-VII discusses the prediction of queue lengths for future time steps using DMD-based algorithms. Finally, section-VIII compare the prediction result of DMDc with LSTM and discusses the implications in the broader context.
\section{Background and Related Works}\label{sec:LR}

The research community has a recent interest in exploiting Koopman theory and related DMD-based algorithms for understanding complex dynamic systems. The eigendecomposition of the Koopman operator can provide an in-depth insight into the system behavior. The applications of Kooman theoretic formalism can be classified into three major categories,  (i) system identification, (ii) state estimation and prediction, (iii) control \cite{kutz2016dynamic,doi:10.2514/1.J052858}.

The usefulness of DMD, its sister algorithms, and Koopman operator formalism has been recently showcased in nonlinear systems like fluid mechanics, epidemiology \cite{proctor2016dynamic}, neuroscience \cite{brunton2016extracting}, financial trend analysis \cite{mann2016dynamic}, electrical power system oscillation analysis \cite{barocio2014dynamic}, traffic flow analysis \cite{avila2020data}, and so on. The DMD applications in traffic system analysis are still in the nascent stages, as described in the following paragraphs.  


It is essential to mention that, in mechanical systems, sensor data is usually sourced from a large number of spatial points for a relatively small temporal observation. It means the data matrix is ``tall and skinny.'' In these cases, DMD performs exceptionally well as a reduced order model (ROM). On the contrary, there are relatively fewer spatial points in traffic flow analysis (spatial sparsity of fixed sensors); but the observation period is relatively longer. This ''fat and short'' nature of traffic data poses some practical challenges in applying DMD approaches. To overcome this challenge, traffic data such as speed, density, flow, and queue lengths are arranged in a Hankel matrix structure. This variant of DMD is commonly known as Hankel DMD or HDMD. It has been established that application of DMD to Hankel data matrices yields the true Koopman eigenfunctions and eigenvalues with an added advantage of guaranteed convergence
 \cite{arbabi2017ergodic}.

One of the first applications of Hankel DMD in transportation research was demonstrated in \cite{avila2017applications} and  \cite{avila2020data}. Avila and et. al. applied the Hankel DMD algorithm to compute the Koopman modes of freeway traffic flow. These Koopman modes uncovered some intriguing patterns, and the interpretations of the patterns were consistent with practical observations. Furthermore, the Koopman analysis was also explored in traffic flow prediction. HDMD technique can be further extended by adding control, which is commonly known as HDMDc. Esther Ling et. al. \cite{ling2018operator,ling2018koopman} demonstrated the application HDMDc in a signalized traffic intersection. The Hankel DMDc technique was used to detect queue length instability. The authors also developed an instability detection algorithm by exploring computed eigenvalues. In recent times, delay embedding, and Koopman theory was used to decompose a chaotic model into a linear model \cite{brunton2017chaos}.

Traffic state prediction is often performed using time series modeling. Auto‐regressive integrated moving average (ARIMA) models \cite{kumar2015short}, Kohonen autoregressive integrated moving average (KARIMA) method \cite{van1996combining}, Space–time autoregressive integrated moving average (STARIMA) \cite{min2011real} method, different neural network models \cite{kumar2013short}, reinforcement learning \cite{walraven2016traffic} etc. have been explored for traffic prediction. Long short-term memory (LSTM) is also widely used for various problems related to traffic prediction \cite{altche2017lstm,hochreiter1997long,8288312} --- for instance, travel time prediction \cite{7795686}, traffic flow prediction \cite{8317872}, traffic queue prediction \cite{wang2020lane} etc. LSTM is a type of recurrent neural network (RNN), which has been found useful for time-series prediction with high accuracy \cite{gers2000learning}. 

\section{Mathematical Preliminaries}\label{sec:MB}
This section provides the necessary background on the DMD, DMDc, and HDMDc algorithms.

\subsection{DMD and HDMD} \label{sec:DMD}
Let us consider a dynamic system:

\begin{equation}
 x_{k+1} = \textbf{F}(x_k) \label{eq:1}
\end{equation}

\noindent where the discrete-time flow is described by \textbf{F}, $x_k$ denotes the state vector of the dynamics at $k^{th}$ time frame and $x_k \in \mathbb{R}^n$. Now we consider two large data matrices which are comprised of $m$ snapshots each from the same dynamic system. 
\[X=
\begin{bmatrix}
\mid & \mid & & \mid\\
x_{1} & x_{2} & ... & x_{m}\\
\mid & \mid & & \mid\\
\end{bmatrix}
 \mbox {and}\
X'=\begin{bmatrix}
\mid & \mid & & \mid\\
x_{2} & x_{3} & ... & x_{m+1}\\
\mid & \mid & & \mid\\
\end{bmatrix}\\
\]
$$\mbox{where}, \ X, X' \in \mathbb{R}^{n \times m}$$
Considering the locally linear approximation of equation \eqref{eq:1} we can write:

\begin{equation}
 X' = AX \label{eq:2}
\end{equation}

\noindent where $A$ is the best fitting operator which minimizes Frobenius norm of \eqref{eq:2}.
Taking Moore-Penrose pseudoinverse of $X$, equation \eqref{eq:2} can be written as:

 \begin{equation}
A  = X X^ \dagger \label{eq:3}
  \end{equation}

\noindent By applying SVD (singular value decomposition) on $X,$ we get $X=U \Sigma V^*$. Substituting $X$ in equation \eqref{eq:3}

\begin{equation}
A  = X' ({U \Sigma V^*})^{-1} 
    = X'V \Sigma^{-1} U^*  \label{eq:4} 
  \end{equation}

\noindent Now, we can make the computation more efficient by reducing the rank of SVD. Taking the first $r$ dimensions of $U,V$ and $\Sigma$ we get, $U_r,V_r$ and $\Sigma_r$. Hence, A of equation \eqref{eq:4} becomes $A_r$

\begin{equation}
A_r  = X'V_r \Sigma_r^{-1} U_r^*  \label{eq:4_1} 
  \end{equation}

\noindent The computation can be made further efficient by projecting $A_r$ on its POD (proper orthogonal mode) as follows:

\begin{equation}
\Tilde{A}  = U_rA_rU_r^*  \label{eq:4_2}
           = U_rX'V_r \Sigma_r^{-1}
  \end{equation}

\noindent For highly nonlinear dynamical systems, locally linear approximation works well when data matrix $X$ is sufficiently tall and skinny. Otherwise locally linear approximation sometimes does not fit well. This problem is often tackled by using Hankel matrix structure, where time-shifted snapshots are vertically stacked  to create a taller and skinnier matrix. This procedure is also known as time delay embedding. After embedding $h$ snapshots in $X$ and $X'$, Hankel matrix form of the two matrices $\Tilde{X}$ and $\Tilde{X'}$ can be expressed as follows:
\begin{equation*}
\Tilde{X}=\begin{pmatrix}
 \mid & \mid &   & \mid\\
x_{1} & x_{2} & ... & x_{m-h}\\
\mid & \mid &   & \mid\\
x_{2} & x_{3} & ... & x_{m-h+1}\\
\mid & \mid &    & \mid\\
. & . &    & .\\
. & . &    & .\\
\mid & \mid &   & \mid\\
x_{h} & x_{r+1} &... & x_{m}\\
\mid & \mid &    & \mid\\
\end{pmatrix}
\\ \mbox{and} 
\end{equation*}

\begin{equation*}
\Tilde{X'}=\begin{pmatrix}
 \mid & \mid &   & \mid\\
x_{2} & x_{2} & ... & x_{m-h+1}\\
\mid & \mid &   & \mid\\
x_{3} & x_{3} & ... & x_{m-h+2}\\
\mid & \mid &    & \mid\\
. & . &    & .\\
. & . &    & .\\
\mid & \mid &   & \mid\\
x_{h+1} & x_{r+1} &... & x_{m+1}\\
\mid & \mid &    & \mid\\
\end{pmatrix}\\
\end{equation*}
\mbox{where}, $\Tilde{X}, \Tilde{X'} \in \mathbb{R}^{hn \times (m-h)}$

\subsection{DMDc and HDMDc}\label{sec:HDMDC}
We can incorporate control inputs to DMD by considering a modified (DMDc) algorithm. The main goal of DMDc is build up a relationship between the present state $x_k$, a future state $x_{k+1}$ and the control $u_k$ \cite{kutz2016dynamic}. The canonical form of the dynamical system can be written as:
  
  \begin{equation}\label{eq:5}
     {x}_{k+1}  = A x_k +B u_k
  \end{equation}
  
\noindent where, $x_k,x_{k+1} \in \mathbb{R}^n$.  
Equation \eqref{eq:5} can be written in matrix form  as:

\begin{equation}\label{eq:6}
X'  = A X +B \Upsilon
  \end{equation}
  
\noindent With a simple matrix manipulation equation \eqref{eq:6} can be rewritten as:

\begin{equation}
  X' = \begin{bmatrix}
       A & B
      \end{bmatrix} \begin{bmatrix}
         X\\
       \Upsilon
   \end{bmatrix}=  G\Omega
\end{equation}
  
\noindent where, the sequence of control action if given by 
\begin{equation*}
\Upsilon =
\begin{bmatrix}
\mid & \mid & & \mid\\
u_{1} & u_{2} & ... & u_{m}\\
\mid & \mid & & \mid\\
\end{bmatrix}
\end{equation*}
Here, $ X \in \mathbb R^{n \times m}$, $X'\in\mathbb R^{n \times m}$,  $u_k\in\mathbb R^{q}$, $A \in \mathbb R^{n \times n}$, $B \in \mathbb R^{n \times q}$, $G \in \mathbb R^{n \times (n+q)}$, and $\Omega \in \mathbb R^{n \times (n+q)}$.

Now, we consider $h$ time delay embedding of $X$ and $X'$ thus obtaining $\Tilde{X}$ and $\Tilde{X}'$. $G$ becomes $\Tilde{G}$. Instead of DMDc we will refer to the model as Hankel DMDc or HDHDc. Taking SVD of $\Tilde{X} = {\Tilde{U} \Tilde{\Sigma} \Tilde{V}^*}$ and applying the arguments of equation \eqref{eq:3} and equation \eqref{eq:4} we obtain: 
   \begin{equation}
    \Tilde{G} = \Tilde{X'} ({\Tilde{U} \Tilde{\Sigma} \Tilde{V}^*})^{-1} 
    = \Tilde{X'} \Tilde{V} \Tilde{\Sigma} \Tilde{U}^*   
  \end{equation}
  
Here, $ \Tilde{X} \in \mathbb R^{nh \times (m-h)}, \Tilde{X}' \in \mathbb R^{nh \times (m-h)},  \Tilde{U} \in \mathbb R^{(n+q)h \times (m-h)}$. By spliting $\Tilde{U}$ vertically in two components $ \Tilde{U_1} \in \mathbb R^{nh \times (m-h)}$ and $\Tilde{U_2} \in \mathbb R^{qh \times (m-h)}$  
we can write $\Tilde{U}$  =  $$\begin{bmatrix}
\Tilde{U_1}\\
\Tilde{U_2}
\end{bmatrix}$$
Now we can compute $\Tilde{A}$ and $\Tilde{B}$ 
by using the following equations:

 \begin{equation}
\Tilde{A}= \Tilde{X'V} \Sigma^{-1} \Tilde{U_1}^*
\end{equation}
\begin{equation}
\Tilde{B}= \Tilde{X'V} \Sigma^{-1}\Tilde{U_2}^*
\end{equation}
  
$\Tilde{A}$ and $\Tilde{B}$ can be used for various applications such as future state prediction, system identification, designing control algorithm, etc. In general, we can reduce the dimension of both $\Tilde{A}$ and $\Tilde{B}$ by applying the same procedures as of equation \eqref{eq:4_1} and equation \eqref{eq:4_2}. Note that there is no accepted approach for choosing the data snapshots used for training $\Tilde{A}$ and $\Tilde{B}$, which is an open research question. 

\section{Data Description }
The signal timing data for the intersections in this study was collected from the signal retiming report \cite{kevinr.carey2017} prepared by Faller, Davis \& Associates Inc. (FDA) for Florida Department of Transportation (FDOT) district $5$. The signal retiming information of S.R. $50$ at Murdock Boulevard and S.R. $50$ at Rouse Road located in Orlando, Orange County, Florida, were used for simulation in this study. {Both the signals had fixed signal timings.} The study area is represented in Fig. \ref{fig:1}:

\begin{figure}[H]
\centering
  \begin{subfigure}{.24\textwidth}
    \includegraphics[width=1\textwidth,height=\textwidth]{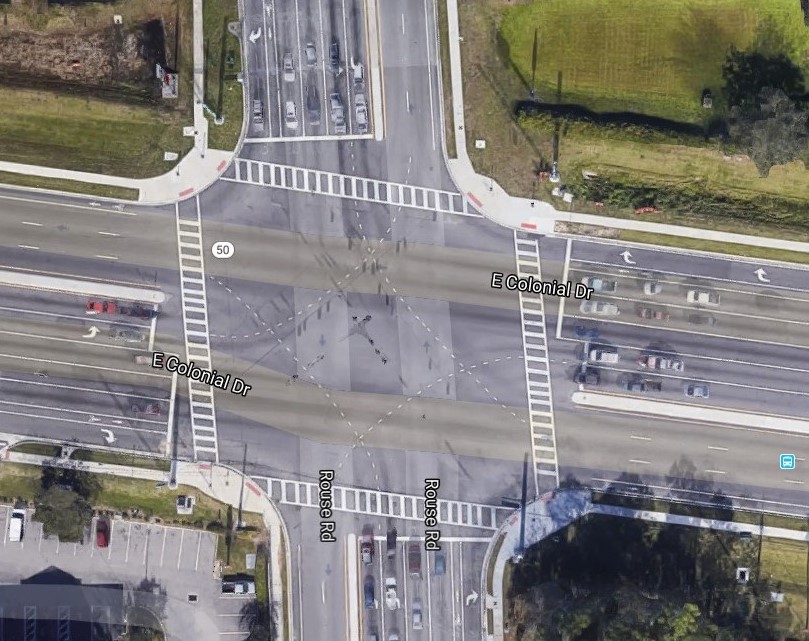}
    \caption{Rouse Intersection }
    \label{fig:1a}
  \end{subfigure}%
  ~
  \begin{subfigure}{.24\textwidth}
    \includegraphics[width=1\textwidth,height=\textwidth]{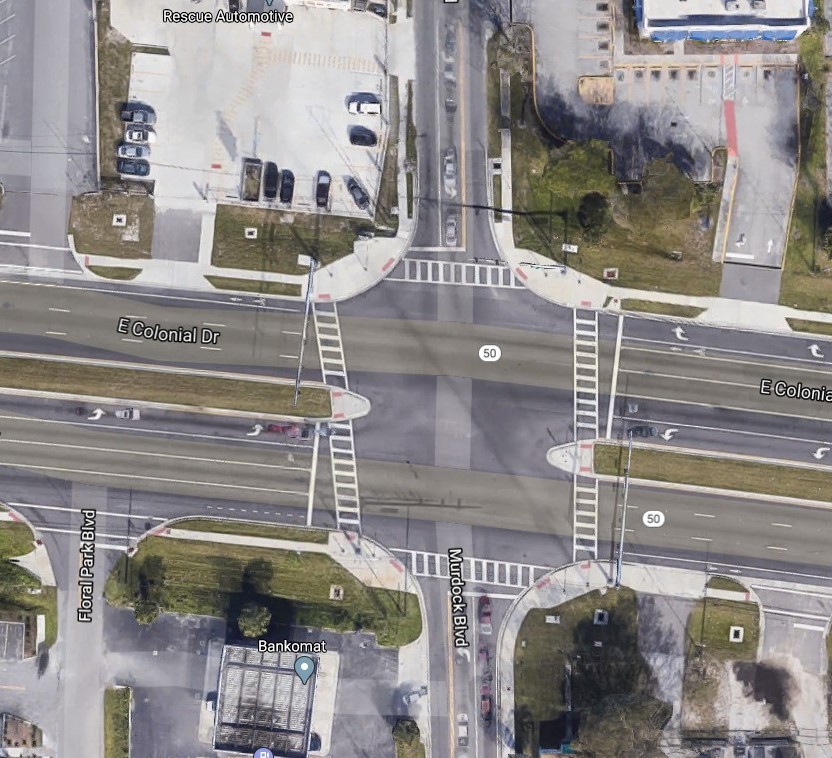}
    \caption{Murdock Blvd. Intersection}
    \label{fig:1b}
  \end{subfigure}
  \caption{Satellite images of the two intersections collected from google map.}
  \label{fig:1}
  \centering
\end{figure}

The vehicle turning movement count aggregated for 15 minutes was collected from signal timing data for the intersections S.R. $50$ at Murdock Boulevard and S.R. 50 at Rouse Road used. The vehicle movement data was collected from $8.00$ a.m. to $9.00$ a.m. on February, $17^{th},2017$. The vehicle's average speed in the network was collected from radar detectors of the Regional Integrated Transportation Information System (RITIS) website \cite{RITIS}.

\section{Simulation Model Development and Calibration}
Microscopic traffic simulation is used to simulate traffic scenarios in a computer-based environment \cite{barcelo2005microscopic}. There are different traffic simulation software, for example, VISSIM, PARAMICS, CORSIM, SUMO, etc. In this study, SUMO which is open-source software was used for developing the simulation model. SUMO has a unique graphical interface to visualize scenarios that can handle large networks \cite{Krajzewicz2010}.

The U.S. Department of transportation provided guidelines of micro-simulation for efficiently replicating traffic operation theory \cite{dowling2004traffic}. The simulation model development and model calibration were performed following the U.S. Department of Transportation guidelines. The simulation model was finally validated with real-world traffic data.

The network for performing the simulation was extracted from overpass turbo, an open street map API. The simulation was run in SUMO from $8.00$ a.m. to $9.00$ a.m. in the study area. The simulation's first and last fifteen minutes were considered a warm-up and cool-down periods (no data were extracted during this time window). For calibration and validation simulation, results from $8.15$ a.m. to $8.45$ a.m. was considered.  The signal timing data was incorporated in the network file before simulating SUMO. The route file was prepared from the vehicular movement data aggregated for $15$ minutes in all directions (northbound, southbound, eastbound, and westbound) along the corridor for the two selected intersections. The vehicular movement along each direction was of five types, namely: u-turn, left turn, through, right turn, and right turn on red. All these movement patterns were included in the simulation.

\begin{figure}[H] 
\centering
  \begin{subfigure}{.24\textwidth}
    \includegraphics[width=1\textwidth]{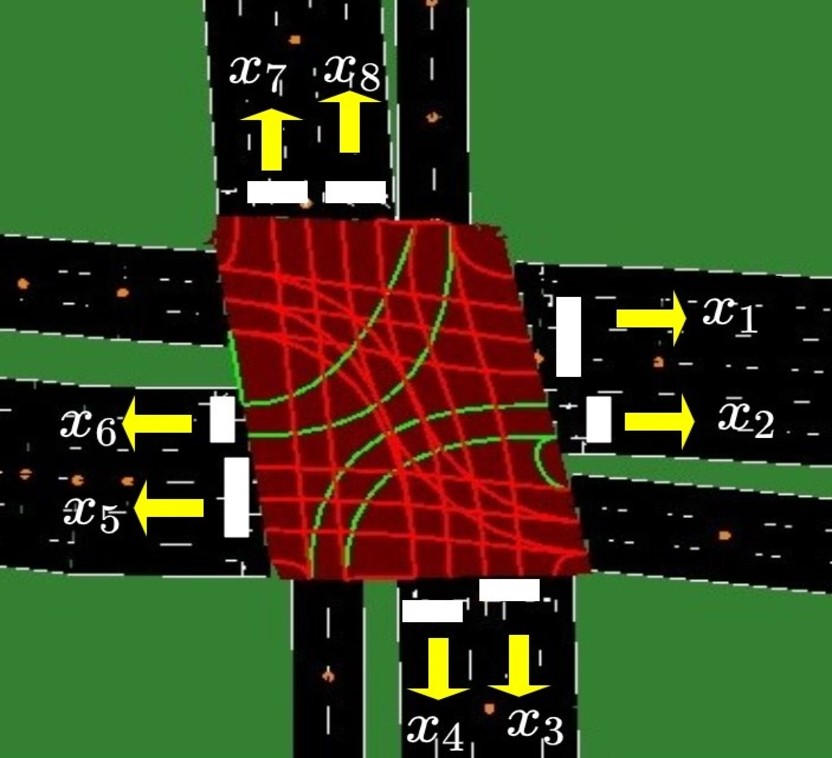}
    \caption{Rouse Intersection }
    \label{fig:2a}
  \end{subfigure}%
  ~
  \begin{subfigure}{.24\textwidth}
    \includegraphics[width=1\textwidth]{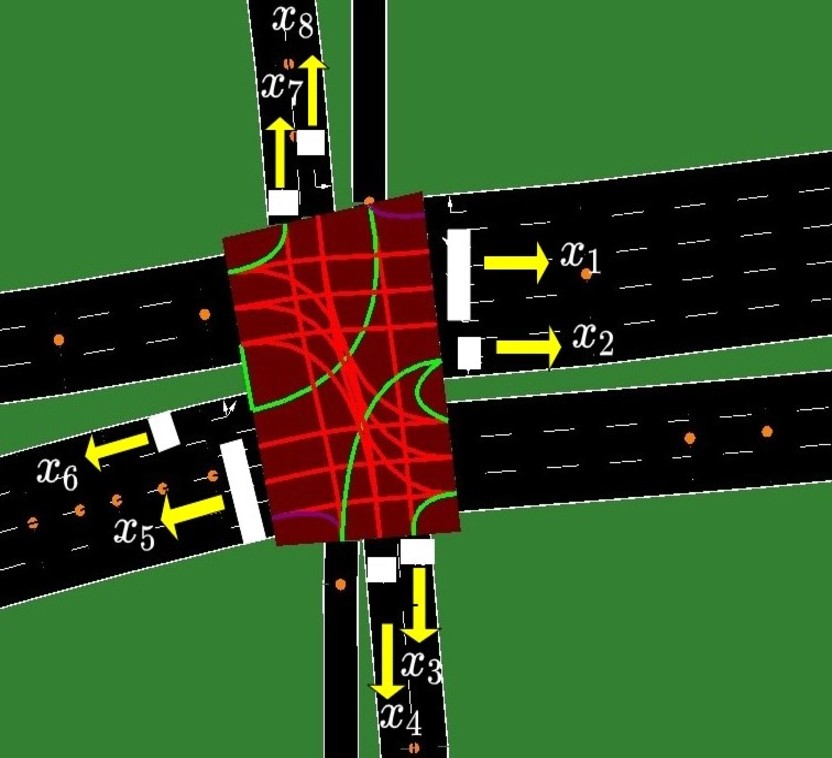}
    \caption{Murdock Blvd. Intersection}
    \label{fig:2b}
  \end{subfigure}
  \caption{Traffic states for vehicle turning movement at Rouse intersection and Murdock Blvd. intersection displayed in SUMO netedit}
  \label{fig:2}
  \centering
\end{figure}

Traffic volume was used for calibrating the parameters in this study. In this study, Geoffrey E. Heavers (GEH) statistics were used for model calibration. This considers both the percentage difference and absolute value. GEH defines the goodness of fit of the model \cite{feldman2012geh}. The equation for the calculation of GEH is as follows:

  \begin{equation}
    GEH = \sqrt\frac{2\times(V_{obs}-V_{sim})^2}{(V_{obs}+V_{sim})}\\
  \end{equation}

In the above equation, $V_{obs}$ are the traffic volumes travelling in different directions and aggregated for $15$ minutes at each detector near the intersection. $V_{sim}$ is the traffic volumes of the same $8$ detectors in the simulation. The simulation model is considered as a good fit if the value of GEH is less than $4$ for sum of vehicles of all the links\cite{dowling2004traffic,nezamuddin2011technical}. The value of GEH for the calibrated parameters was $0.78$ for this study. This value shows that the simulated volume of vehicles replicate the real field volume. The calibrated values are displayed in the table below:

\begin{table}[H]
	\caption{SUMO calibration parameters}
	\begin{center}
		\begin{tabular}{|l|l|l|l|l|}
		\hline
			\textbf{Parameters} & \textbf{Unit} & \textbf{Default value} & \textbf{Range} & \textbf{Calibrated value}\\\hline
			Acceleration   & $m/s^2$ & $2.6$ & $2.6-3.6$ & $3.4$ \\\hline
			Deceleration   & $m/s^2$ & $4.5$ & $4.5-5.5$ & $4.5$\\\hline
			Tau & N/A & $1$ & $1-1.5$ & $1.5$\\\hline
			Sigma & N/A & $.5$ & $.1-.5$ & $.3$\\\hline
		\end{tabular}
	\end{center}
	\label{tab 1}
\end{table}
The four parameters in Table \ref{tab 1} were adjusted for model calibration. The first and second parameters are acceleration and deceleration ability of vehicles. The third and fourth parameters are tau (Driver's minimum time headway) and sigma (driver's imperfection). {The driver's imperfection in SUMO is defined by the combined errors of perception, processing, and actuation of driver's maneuvers.}

Moreover, for calibrating the traffic simulation model, the Correlation Coefficient (CC) was calculated, representing a linear correlation of real-world data and simulated data \cite{el2011calibration,hollander2008principles}. The formula for calculation of correlation coefficient is as follows:

  \begin{equation}
    CC = \frac{\sum(x_{obs}-\Bar{x}_{obs})(x_{sim}-\Bar{x}_{sim})}{\sqrt{\sum(x_{obs}-\Bar{x}_{obs})^2}\sqrt{\sum(x_{obs}-\Bar{x}_{obs})^2}}
  \end{equation}

In the above equation, $x_{obs}$ and $x_{sim}$ are the real world and simulated aggregated traffic volume of each incoming edges towards the two intersections for 15 minutes, respectively. $\Bar{x}_{obs}$ and $\Bar{x}_{sim}$ are the average of real-world traffic volume and the average of simulated traffic volume, respectively. The CC value greater or equal to $.85$ is deemed to be acceptable for the calibrated model. In this study, the value of the Correlation Coefficient was $0.96$, which demonstrates the proper calibration of the parameters.

For validating the simulation model, the average field speed from the detectors was used. {The detectors were considerably far from the intersection.} Concerning speed, the absolute difference between field speed and simulated speed must be inside five miles per hour for $85$ percent of the cases \cite{nezamuddin2011traffic}. In our model, in $100$ percent of the cases, the absolute speed difference between the field average speed and simulated average speed of the detectors was below $5$ miles per hour. This result implies the developed traffic simulation model's validity and infers that the simulated model is consistent with real-world traffic conditions.

Finally, the simulation was run in SUMO for 30 minutes. The queue length output was extracted using python code from the simulation. The queue length data was generated for $1$ sec aggregation. There were 8 incoming edges for the two intersections. We collected the queue length data of the $8$ edges separately. Furthermore, for applying dynamic mode decomposition with control, we also extracted the color of a traffic light for each of the eight edges for every second. These two data sets of queue length and traffic light control parameter in binary form were used to analyze this paper.

\section{System Identification \& Modeling}\label{sec:modeling} 
This section formulates a system identification problem for the signalized intersections. We also provide details on the application of DMDc and HDMDc.

We consider traffic queue dynamics of the Rouse Road and the Murdock Blvd. intersections. There are a total of 8 different turning movements, namely, eastbound, westbound, northbound, southbound, eastbound left, westbound left, northbound left, and southbound left, controlled by signals at each of the intersections. Queue lengths for each movement is considered as a state $x_k$ of the system (see figure \ref{fig:2}). Hence, the state vector of the dynamics is $\mathbf{x} \in \mathbb{R}^8$. 

Traffic lights at the intersections serve as the control input to the queue length dynamics. Therefore, the signal timings for red, green, and yellow lights can be converted to represent control inputs. We have modeled this phenomenon by considering a binary control variable where green and yellow light is treated as $1$ and the red light as $0$. Mathematically it can be presented as follows:

  $u_k(i) =$
  $$\begin{cases}
                                   0 & \text{for red light} \\
                                   1 & \text{for green and yellow light  }
  \end{cases}$$

\begin{figure}[H]
\centering
  \begin{subfigure}[b]{0.24\textwidth} 
    \includegraphics[width=\textwidth]{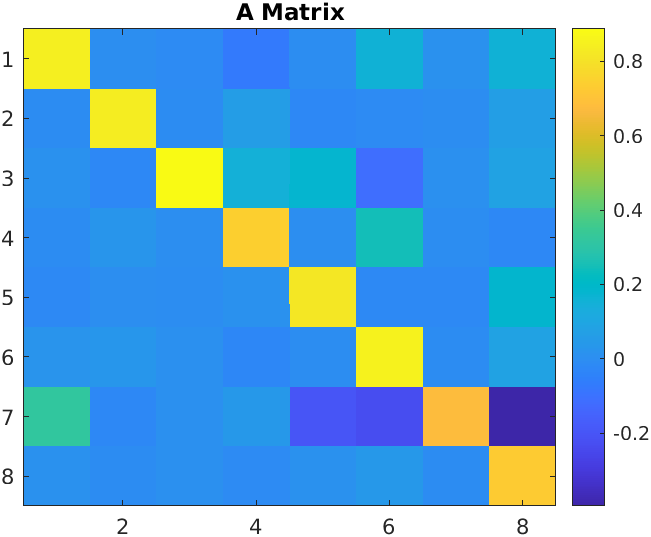}
    \caption{$A$ Matrix for Rouse Road intersection estimated via DMDc}
    \label{fig:3a}
  \end{subfigure}%
  ~
  \begin{subfigure}[b]{.24\textwidth}
    \includegraphics[width=\textwidth]{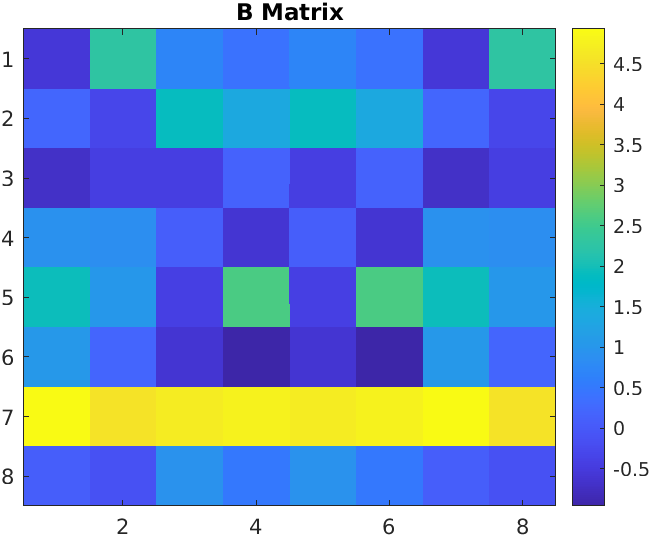}
    \caption{$B$ Matrix for Rouse Road intersection estimated via DMDc}
    \label{fig:3b}
  \end{subfigure}
  ~
  \begin{subfigure}[b]{.24\textwidth} 
    \includegraphics[width=\textwidth]{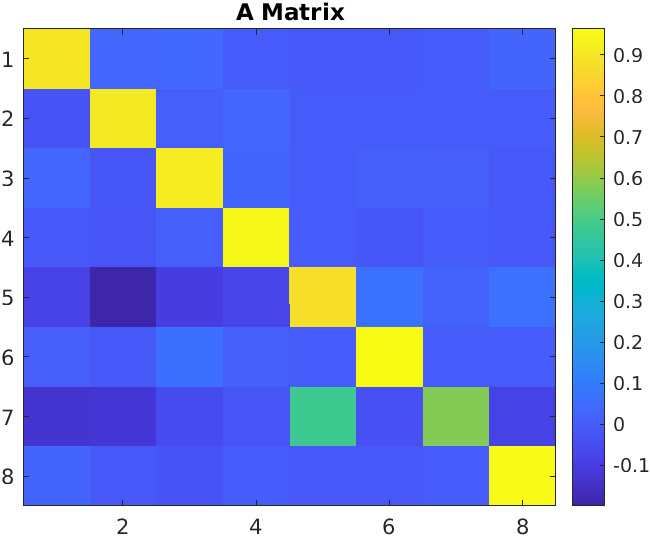}
    \caption{$A$ Matrix for Murdock intersection estimated via DMDc}
    \label{fig:3c}
  \end{subfigure}%
  ~
  \begin{subfigure}[b]{.24\textwidth}
    \includegraphics[width=\textwidth]{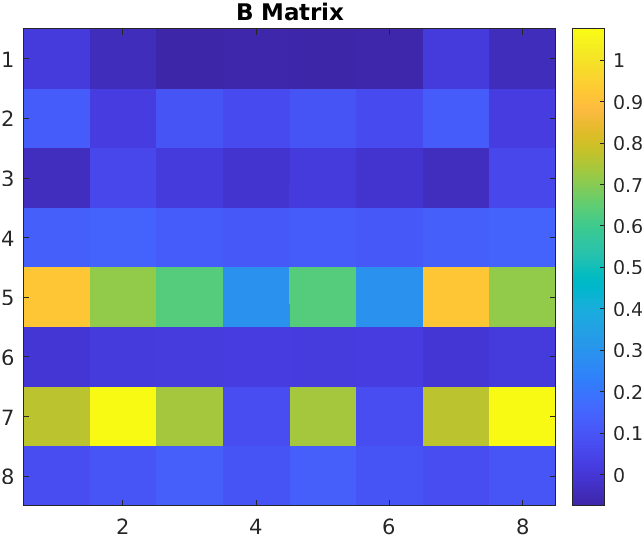}
    \caption{$B$ Matrix for Murdock intersection estimated via DMDc}
    \label{fig:3d}
  \end{subfigure}
\caption{$A$ and $B$ matrices estimated by using DMDc}
\label{fig:3}
\centering
\end{figure}

Recall that DMD does not utilize any physical properties; instead, it captures the best-fitting linear operator using data snapshots for the given dynamical system. The estimation of state matrix $A$ and the control matrix $B$ via the DMDc algorithm mainly depends on the number of snapshots taken for training. However, HDMDc-based system identification also depends on delay-coordinate embedding ($h$) or the Hankel matrix dimension.

\begin{figure}[H]
\centering
  \begin{subfigure}[b]{0.24\textwidth} 
    \includegraphics[width=\textwidth]{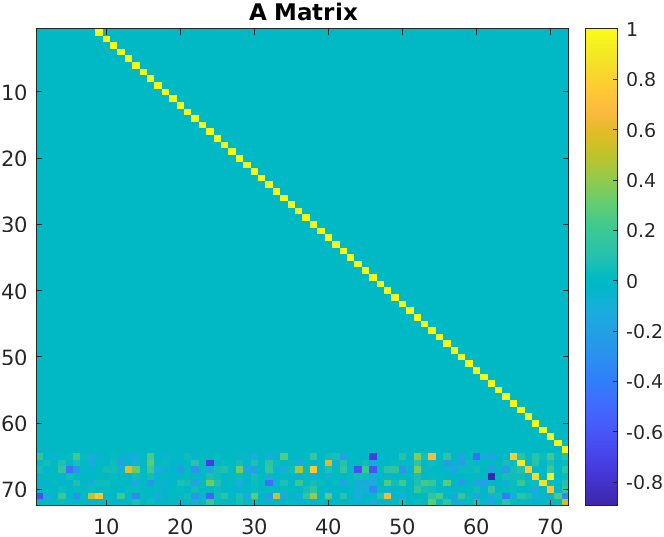}
    \caption{$A$ Matrix for Rouse Road intersection estimated via HDMDc with $h=9$}
    \label{fig:4a}
  \end{subfigure}%
  ~
  \begin{subfigure}[b]{0.24\textwidth}
    \includegraphics[width=\textwidth]{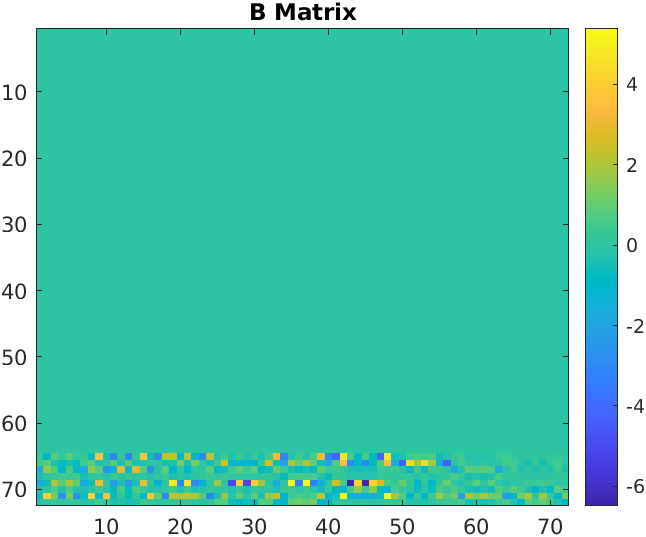}
    \caption{$B$ Matrix for Rouse Road intersection estimated via HDMDc with $h=9$}
    \label{fig:4b}
  \end{subfigure}
  ~
  \begin{subfigure}[b]{0.24\textwidth} 
    \includegraphics[width=\textwidth]{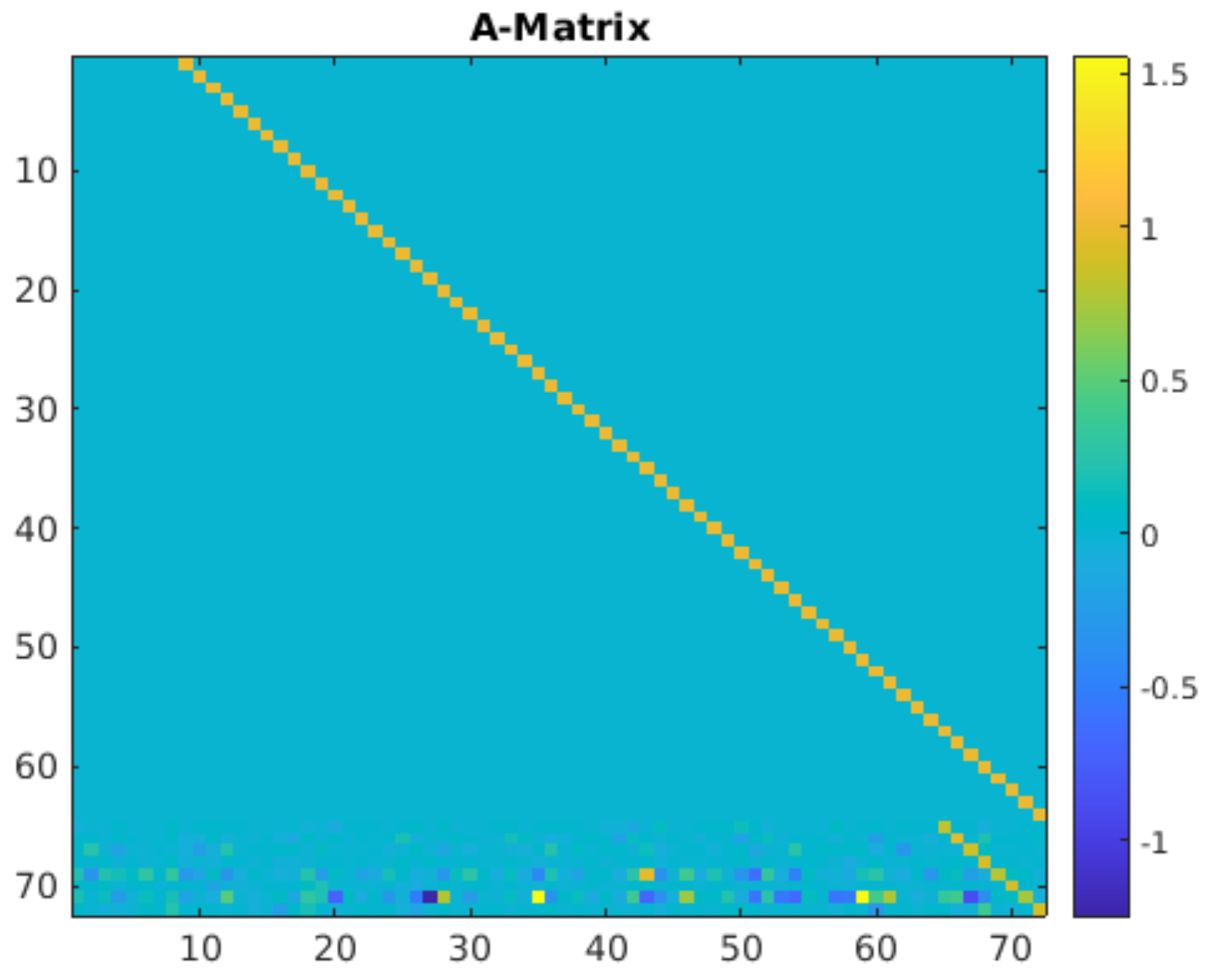}
    \caption{$A$ Matrix for Murdock intersection estimated via HDMDc with $h=9$}
    \label{fig:4c}
  \end{subfigure}%
  ~
  \begin{subfigure}[b]{0.24\textwidth}
    \includegraphics[width=\textwidth]{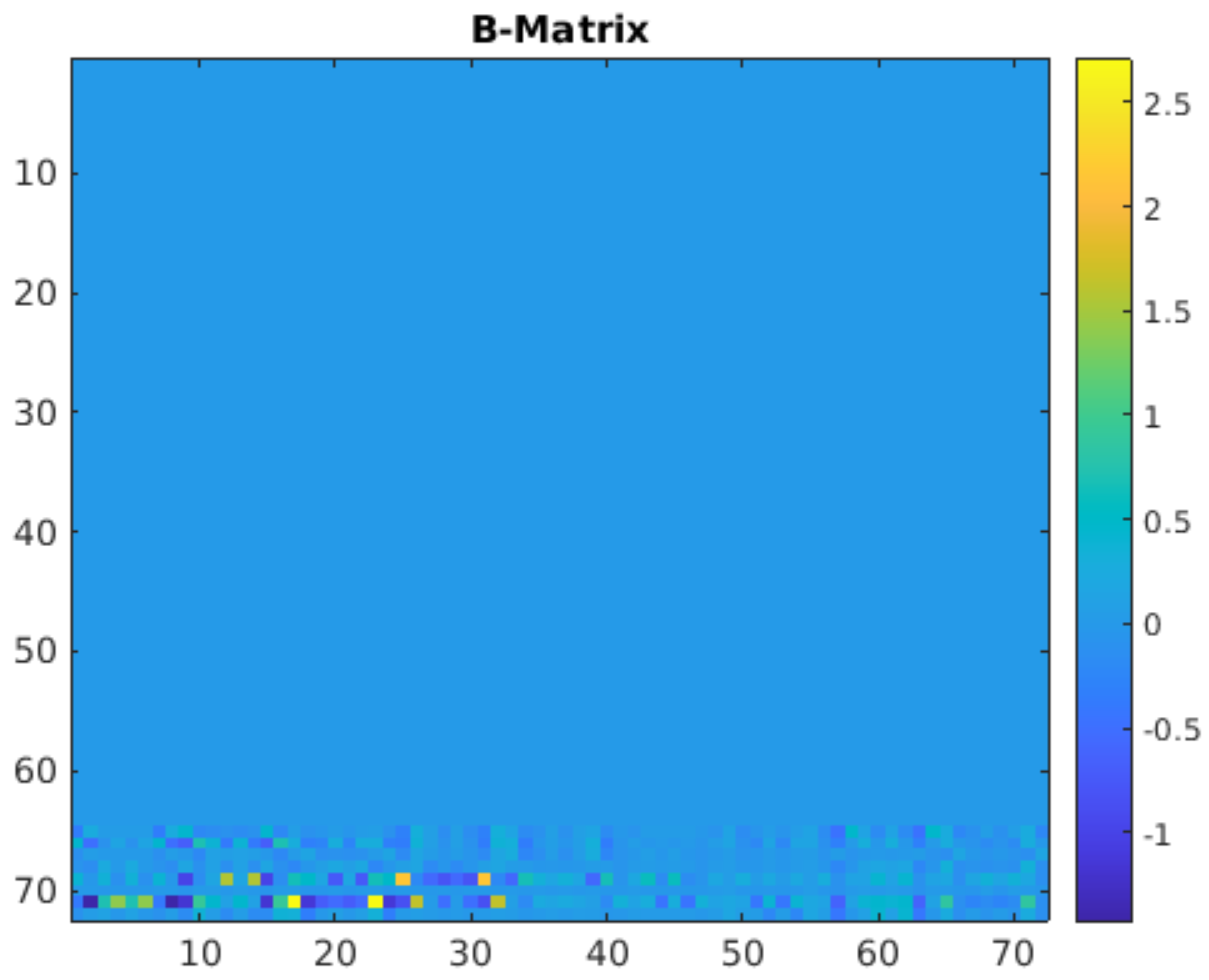}
    \caption{$B$ Matrix for Murdock intersection estimated via HDMDc with $h=9$}
    \label{fig:4d}
  \end{subfigure}
\caption{$A$ and $B$ matrices estimated by using HDMDc}
\centering
\label{fig:4}
\end{figure}

Figure \ref{fig:3} shows the structures of $A$ and $B$ matrices estimated via the DMDc algorithm. We have identified system matrices for the queue dynamics of both intersections. The matrices' elements are displayed as color-coded squares, where $A, B \in R^{8 \times 8}$. On the other hand, the structures of system matrices estimated via the HDMDc algorithm are presented in figure \ref{fig:4}, where $9$ delay-coordinates were embedded. Hence, $A, B \in R^{72 \times 72}$. $400$ snapshots were used for training in each case.

Note that in figure \ref{fig:4}, only the last 8 rows of $A$ and $B$ matrices contain meaningful information (e.g., $A, B \in \mathbb{R}^{8 \times 72}$). Recall that in case of DMDc-based system identification $A,B \in \mathbb{R}^{8 \times 8}$. In HDMDc-based identification, delay-coordinate embedding increases the dimension of system matrices. These extra dimensions of $A$ and $B$ matrices have a physical significance. In DMDc-based system identification, each state depends only on the immediately preceding state. On the contrary, in the HDMDc algorithm embedded with $h$ delay coordinates, each state depends on its preceding $h$ states. This study has chosen $h=9$ or embedded nine delay-coordinates; hence it incorporated additional $64$ columns in the system matrices. These extra columns bridge the current state with preceding $9$ states.

\section{Results and Analysis} 


Using the outcomes of system identification, we perform a wide range of future state predictions. While it is essential to mention that traffic prediction is not a key strength of DMD-based algorithms, the prediction results can quantify the system identification approach's accuracy. This analysis can be valuable in testing the validity of the adopted procedure. Please note that this research aims not to outperform the prediction power of time series methods such as LSTM, ARIMA, or SARIMA. The subsequent comparison with LSTM prediction can only serve as a guide of the DMDc methods' potential. 

In this evaluation, the future states (queue lengths) are estimated using $A$ and $B$ matrices obtained via HDMDc and DMDc algorithms. We used $400$ snapshots for both DMDc and HDMDc-based system identification methods to train the system matrices. Subsequently, we predicted the following $200$ snapshots as a  short-term queue length prediction and $1200$ snapshots as a long-term prediction. 

\begin{figure}[!htb] 
   \centering
  \begin{subfigure}{0.24\textwidth}
    \includegraphics[width=\textwidth]{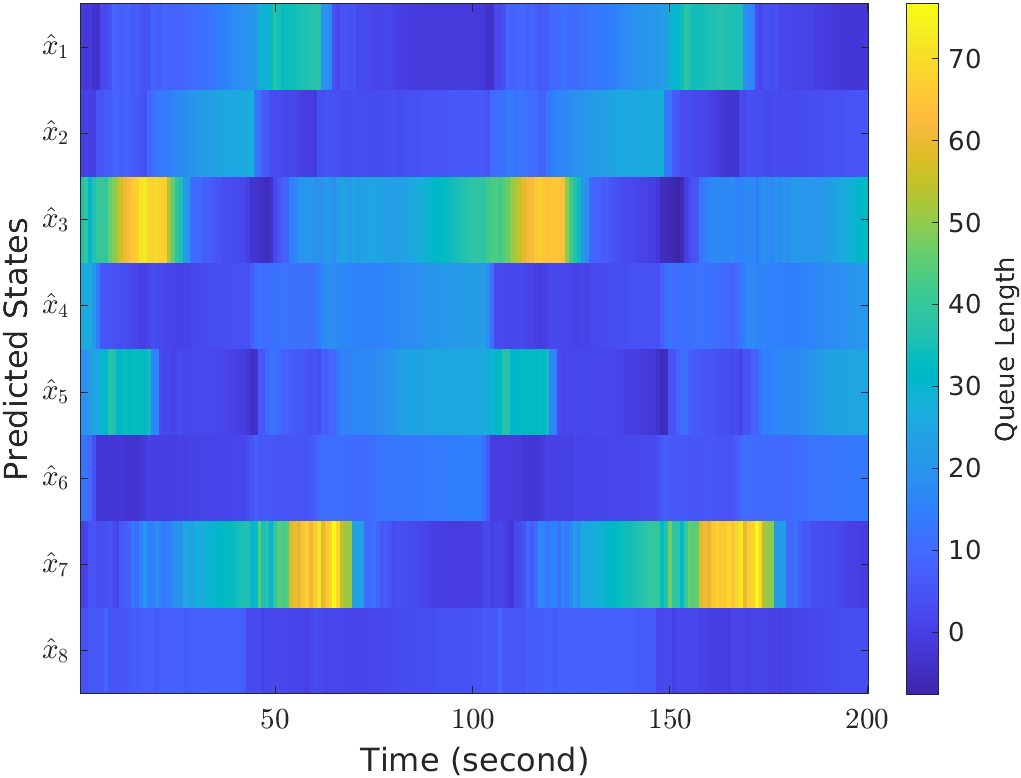}
    \caption{Short-term prediction with HDMDc}
    \label{fig:5a}
  \end{subfigure}%
   ~
  \begin{subfigure}{0.24\textwidth}
    \includegraphics[width=\textwidth]{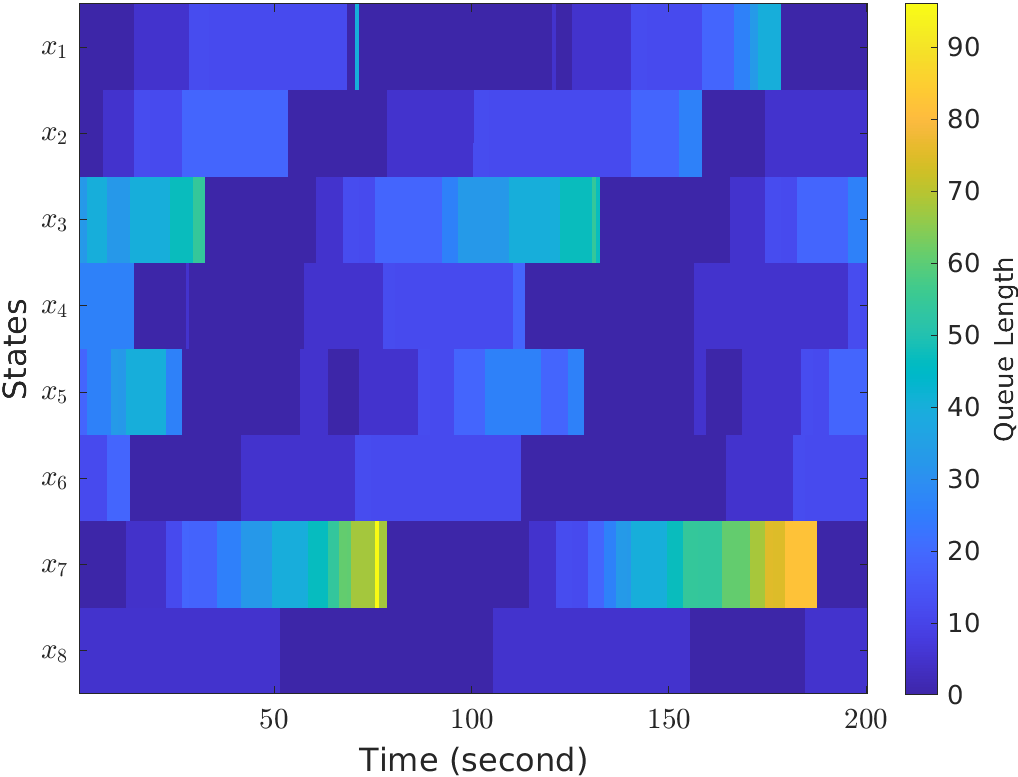}
    \caption{Actual queue length of Rouse Intersection}
    \label{fig:5b}
  \end{subfigure}
   ~
   \begin{subfigure}{0.24\textwidth}
    \includegraphics[width=\textwidth]{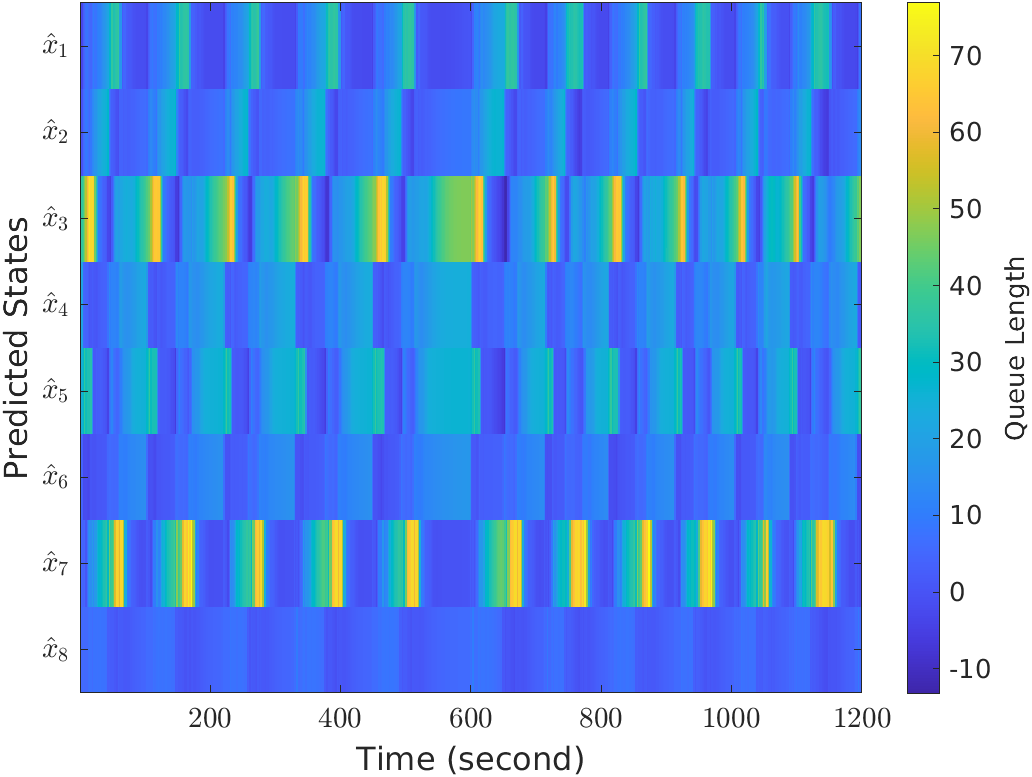}
    \caption{Long-term prediction with HDMDc}
    \label{fig:5c}
    \end{subfigure}%
    ~
   \begin{subfigure}{0.24\textwidth}
    \includegraphics[width=\textwidth]{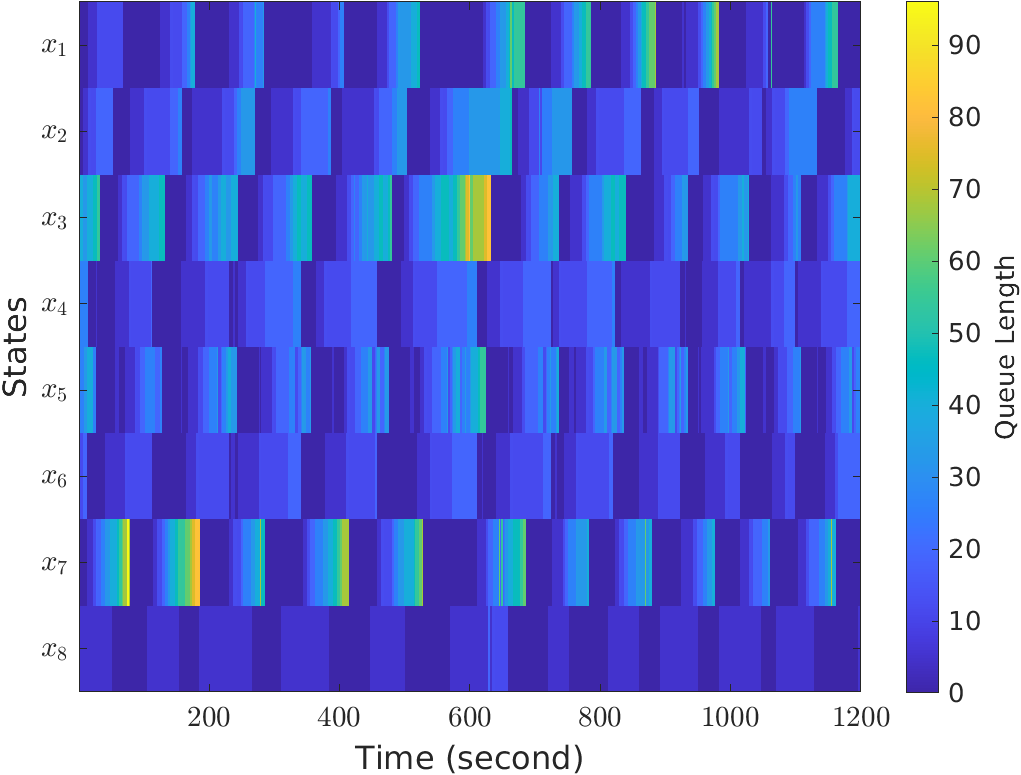}
    \caption{Actual queue length of Rouse Intersection}
    \label{fig:5d}
  \end{subfigure}
   ~  
   \begin{subfigure}[!ht]{0.24\textwidth}
    \includegraphics[width=\textwidth]{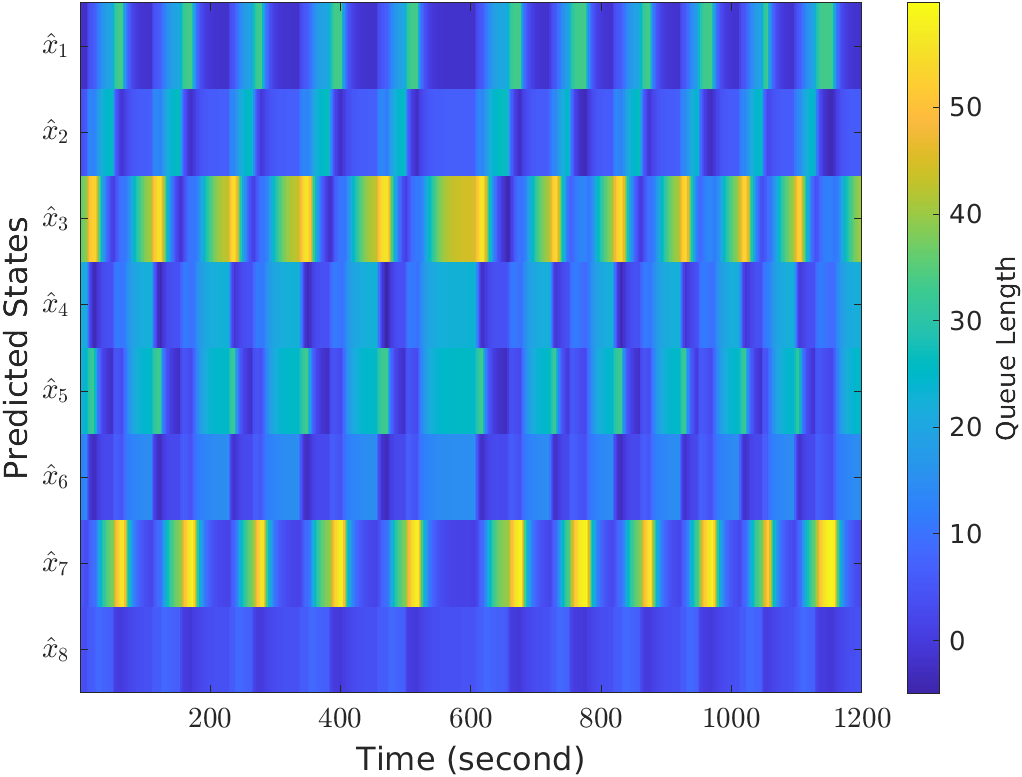}
    \caption{Long-term prediction with DMDc}
    \label{fig:5e}
  \end{subfigure}%
  ~
   \begin{subfigure}{0.24\textwidth}
    \includegraphics[width=\textwidth]{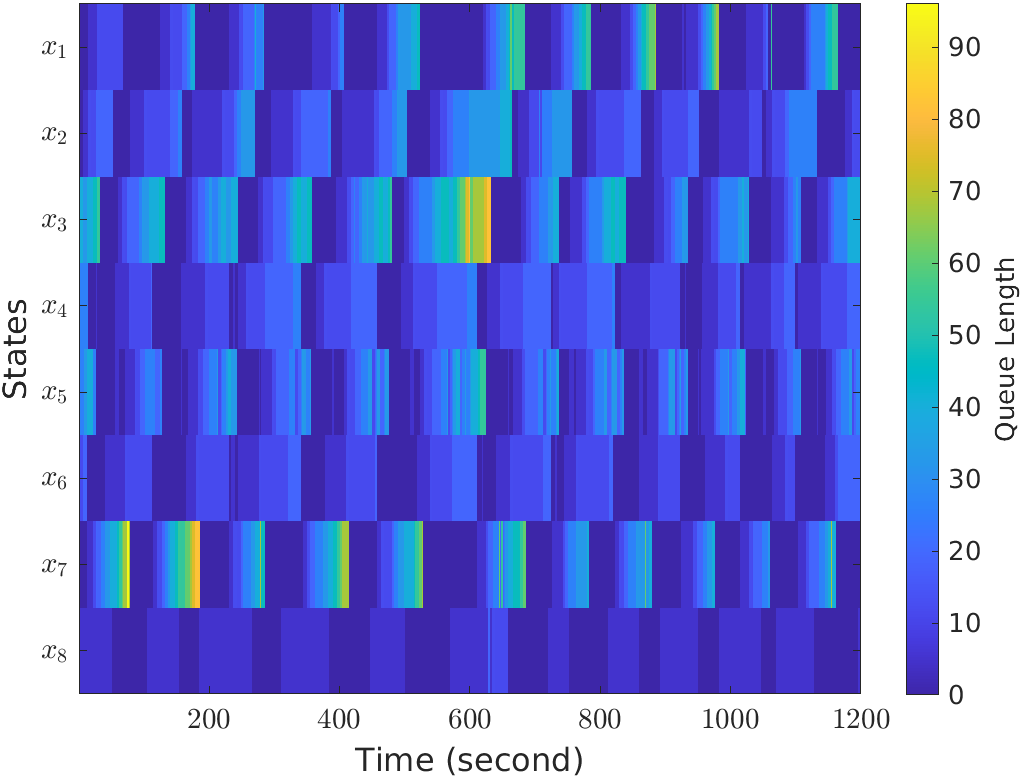}
    \caption{Actual queue length of Rouse Intersection}
    \label{fig:5f}
  \end{subfigure}
   \caption{Prediction results for Rouse road intersection}
   \label{fig:5}
   \centering
\end{figure}

\begin{figure}[H] 
    \centering
  \begin{subfigure}{0.24\textwidth}
    \includegraphics[width=\textwidth]{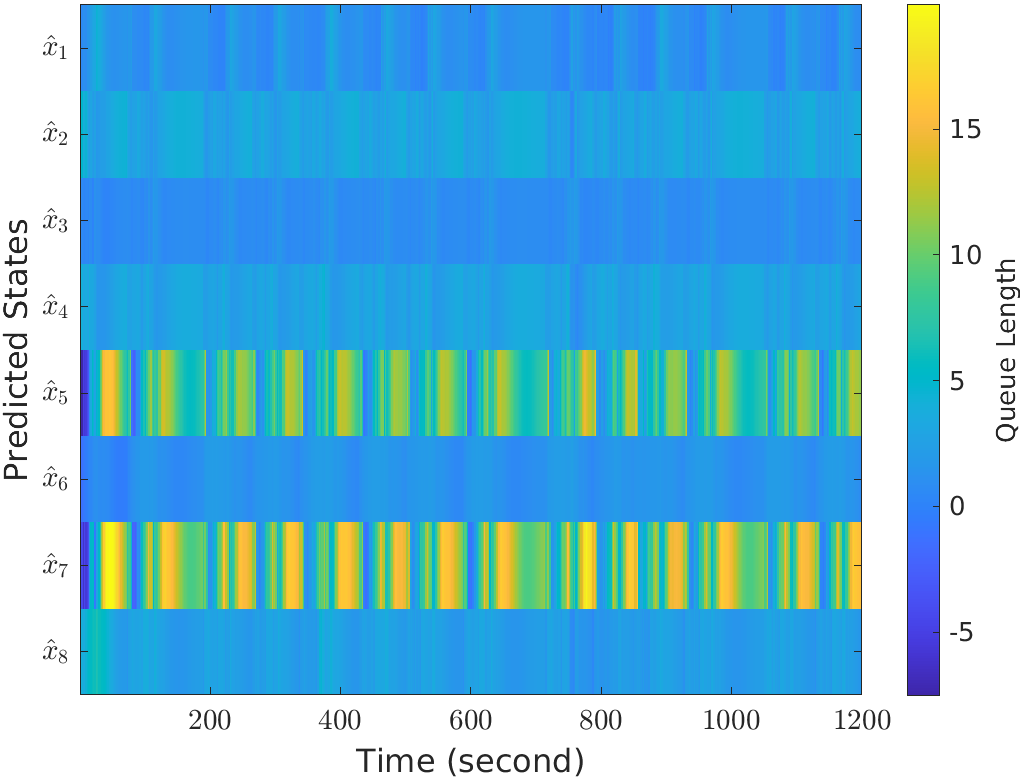}
    \caption{Queue length prediction with HDMDc with $h=9$}
    \label{fig:6a}
  \end{subfigure}%
  ~
  \begin{subfigure}{0.24\textwidth}
    \includegraphics[width=\textwidth]{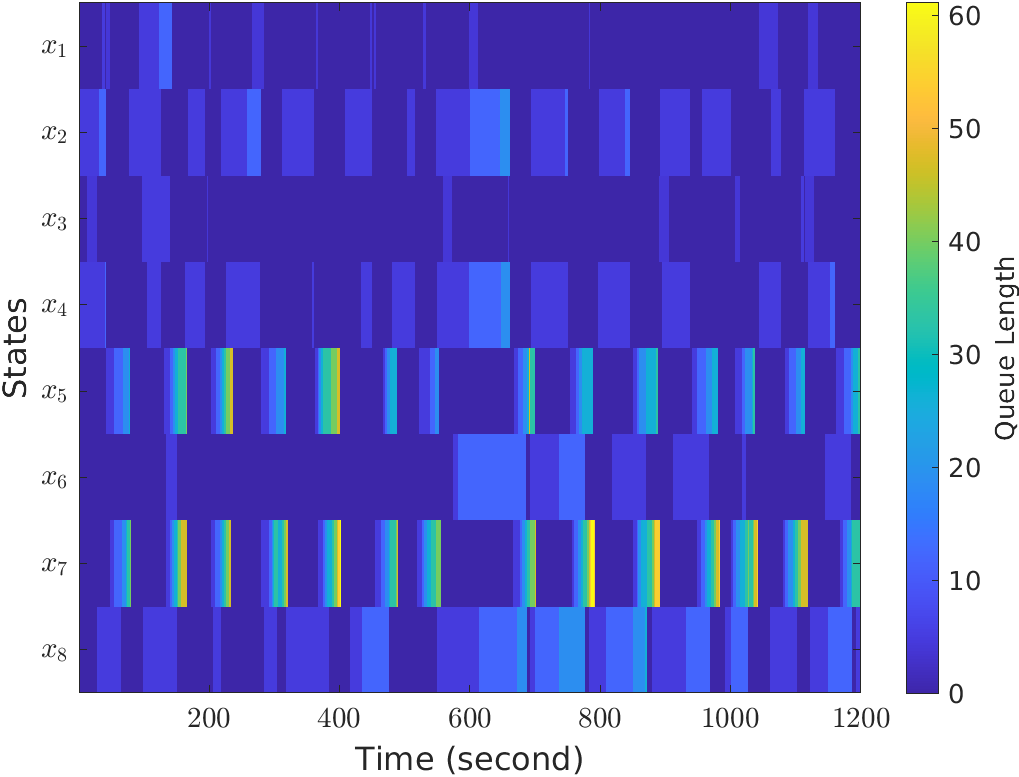}
    \caption{Actual queue length of Murrdock intersection}
    \label{fig:6b}
  \end{subfigure}
   ~
      \begin{subfigure}{0.24\textwidth}
    \includegraphics[width=\textwidth]{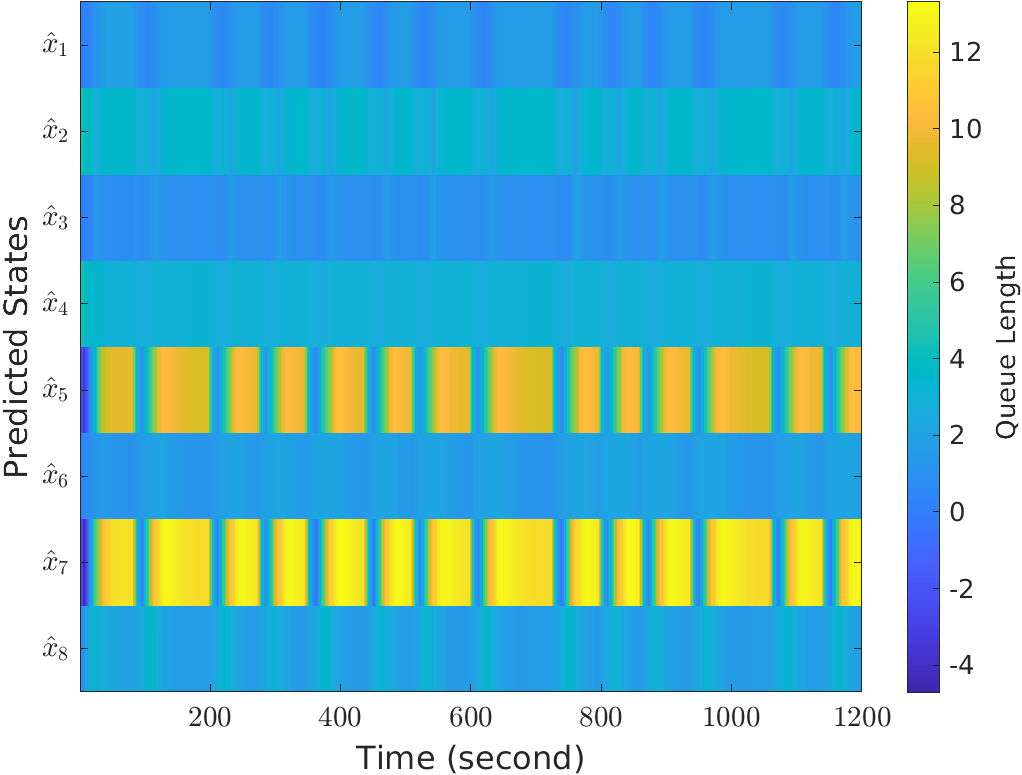}
    \caption{Queue lengths prediction with DMDc}
    \label{fig:6c}
  \end{subfigure}%
    ~
    \begin{subfigure}{0.24\textwidth}
    \includegraphics[width=\textwidth]{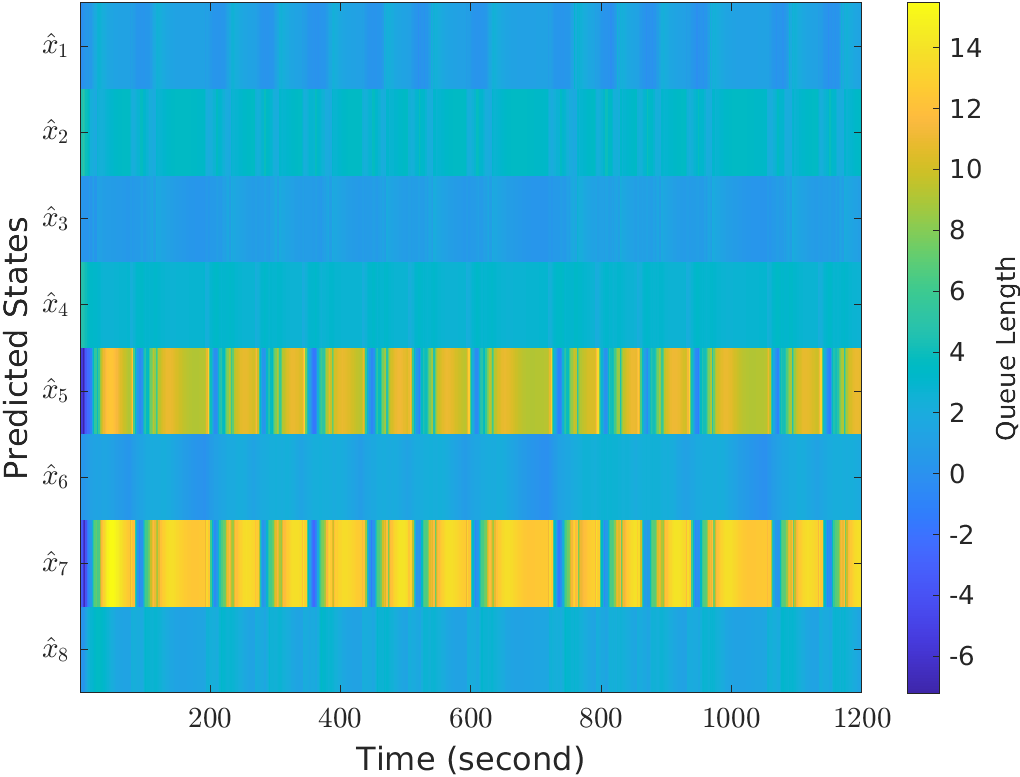}
    \caption{Queue length prediction with HDMDc with $h=5$}
    \label{fig:6d}
    \end{subfigure}
    \caption{Prediction results for  Murdock Blvd intersection}
    \label{fig:6}
    \centering
\end{figure}

The identified dynamical system has been able to predict acceptable results. The predicted queue lengths  at Rouse road intersection are shown in figure \ref{fig:5a}, \ref{fig:5c}, \ref{fig:5e}. Also, The actual queue lengths are shown in figure \ref{fig:5b}, \ref{fig:5d}, \ref{fig:5f}. The predicted queue lengths at the Murdock blvd. intersection are shown in figure \ref{fig:6}. In the figures \ref{fig:5b}, \ref{fig:5d}, \ref{fig:6b},  it is observed that the change of queue states at both  intersections took place with a certain periodicity that can be attributed to the periodic nature of traffic signal control. It is known that DMD and related algorithms can adequately capture the oscillatory behavior present in the data \cite{kutz2016dynamic}. In the predicted  states, the oscillatory patterns of the actual states are preserved {see \ref{fig:5b},\ref{fig:5d},\ref{fig:6b}}.

\subsection{Impact of Delay-Coordinate Embedding}
Recall that the use of delay embedding is necessary, as it makes the data matrix taller and skinnier. 
It is observed that DMDc poorly predicted the sharp increases of queue length (see figure \ref{fig:5e}) in contrast to HDMDc based prediction (see figure \ref{fig:5c}). The result can also be understood from the range of the color bar shown beside the figures. The color bar in figure \ref{fig:5d} shows that the upper limit of queue length in original data is close to $90$, while in HDMDc based prediction shown in figure \ref{fig:5c} it is close to 70. On the other hand, in the DMD based prediction shown in figure \ref{fig:5e} it is just around $50$. That means DMDc could not predict the sharp increases in queue length better than HDMDc. We should be cautious in generalizing these trends. For different intersections, the prediction performances might vary with embedding differently for other data sets.

\begin{figure}[H] 
\centering
  \begin{subfigure}{0.24\textwidth}
    \includegraphics[width=\textwidth]{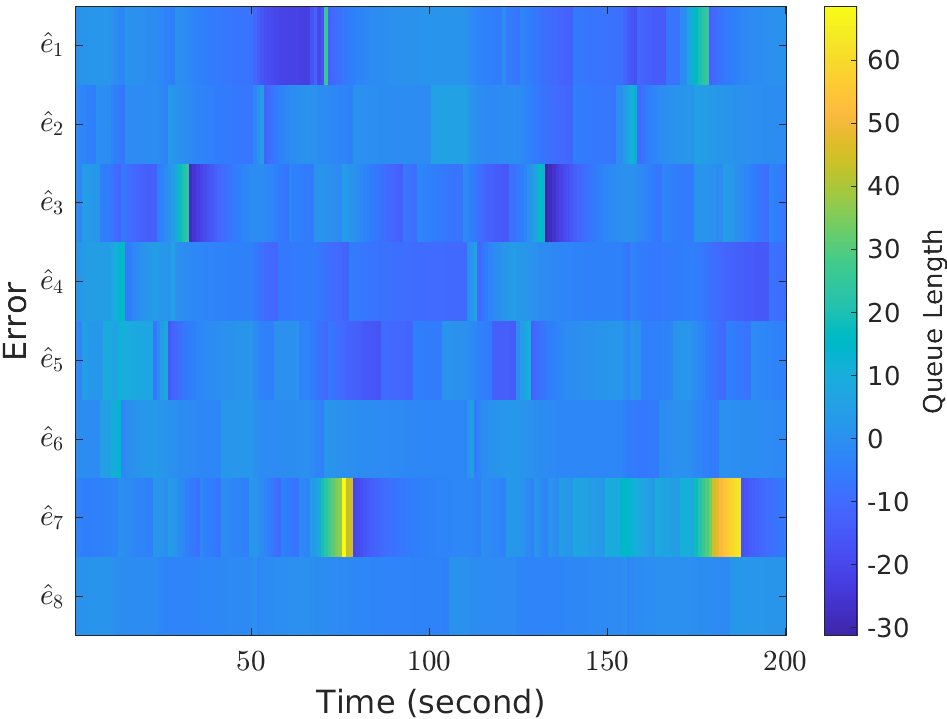}
    \caption{Error for short-term prediction via DMDc }
    \label{fig:7a}
  \end{subfigure}%
  ~
  \begin{subfigure}{0.24\textwidth}
    \includegraphics[width=\textwidth]{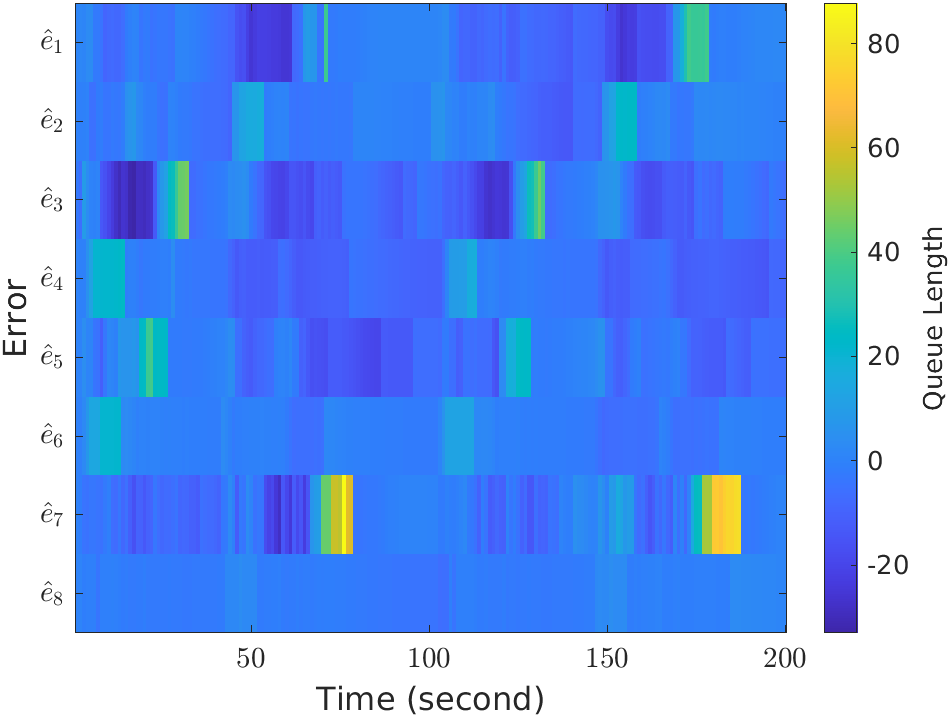}
    \caption{Error for short-term prediction via HDMDc with $h=9$}
    \label{fig:7b}
  \end{subfigure}
  \centering
  ~
  \begin{subfigure}{0.24\textwidth}
    \includegraphics[width=\textwidth]{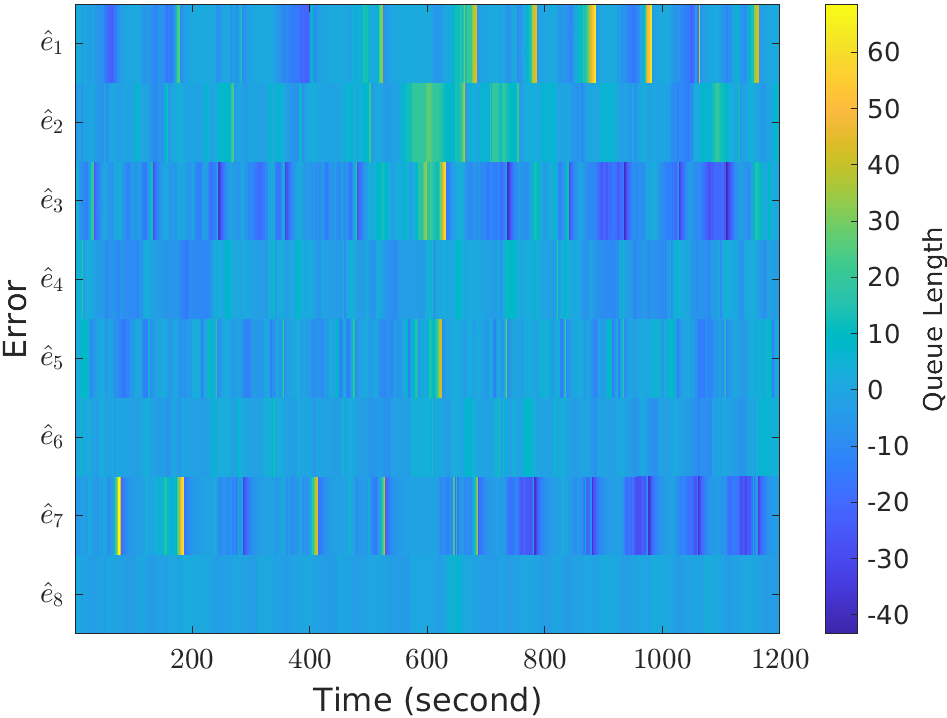}
    \caption{Error for long-term prediction via DMDc}
    \label{fig:7c}
  \end{subfigure}%
  ~
  \begin{subfigure}{0.24\textwidth}
    \includegraphics[width=\textwidth]{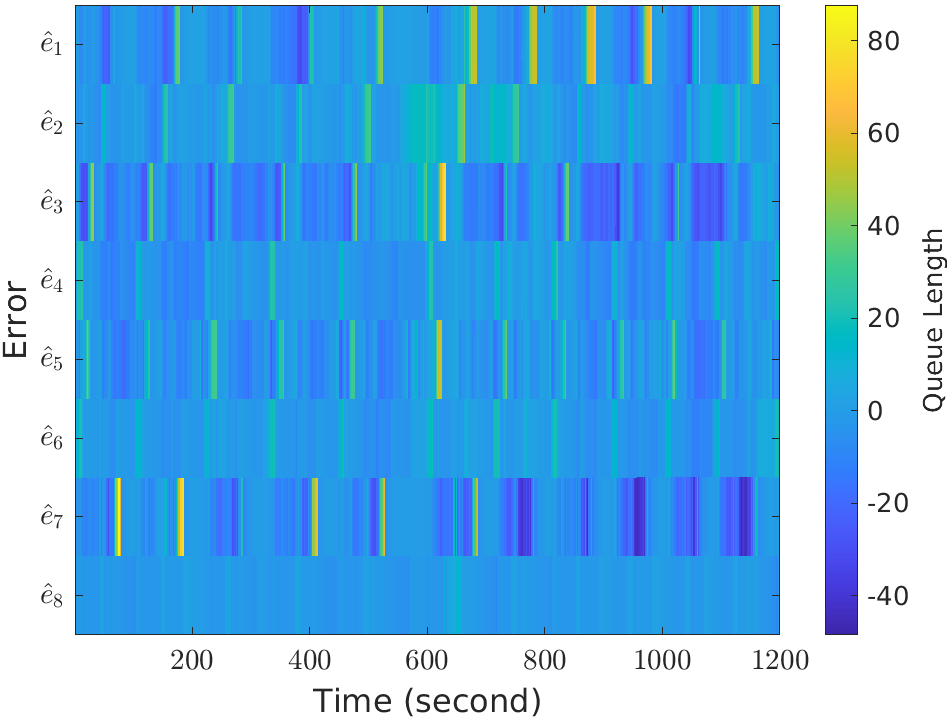}
    \caption{Error for long-term prediction via HDMDc with $h=9$}
    \label{fig:7d}
  \end{subfigure}
  \centering
  \caption{Error $\hat{e}_k = (x_k-\hat{x}_k)$ calculation at different instants at Rouse road intersection}
  \label{fig:7}
\end{figure}

Figure \ref{fig:7} shows the error $\hat{e}_k$ between actual state ($x_k$) and predicted state ($\hat{x}_k$). It can be seen from the figure \ref{fig:7} that for both short-term, and long-term predictions i.e. for $200$s, and $1200$s, DMDc predicts the future states with consistent accuracy. 

As discussed in the previous section, HDMDc based prediction relies on many previous states, whereas in DMDc, future states depend only on the immediately preceding state. As a result, HDMDc is expected to perform more robustly in general. However, the overall prediction performance of HDMDc in our case study was comparable to DMDc, which is shown in figure \ref{fig:8}. One possible explanation may be related to the averaging of errors across all eight states.

The above interpretation of the results underscores a trade-off when choosing the delay embedding. The increase in delay embedding can cause an increase in error. A decrease of the delay embedding, on the other hand, might miss capturing the sharp changes in queue length. As the argument in this research is concerned with HDMDc based system identification, one must consider this dilemma from the two contradicting perspectives while considering the nature of the application at hand.

\subsection{Effects of the Nature of an Intersection}
\noindent We now report on the difference between the nature of two intersections and the impact on the prediction results. From figure \ref{fig:6b}, it is observed that at the Murdock Blvd. intersection, except for two states, all states mainly consist of zero queue length. On the contrary, the Rouse road intersection data is more prosperous and more diverse, as shown in figure \ref{fig:5d}. The Murdock intersection patterns' prediction results were not preserved as accurately as in the Rouse road intersection. This inaccuracy was held even after embedding 9 and 4 delay coordinates, respectively.
The sudden increases could not be predicted as illustrated from the color bars of figure \ref{fig:6a}, \ref{fig:6c}, and \ref{fig:6d}. 


\subsection{Impact of Training Snapshots}
While estimating system matrices with a different number of snapshots, we observed an interesting trend in the choice of training snapshots. Shown in figure \ref{fig:8a} and \ref{fig:8b} are the effect of training snapshots for different value of delay-coordinate embedding $h$ at the Rouse intersection. It is observed that if a fewer number of snapshots are taken, prediction results drastically deteriorate, and the error rises abnormally. In this case, the trend started if less than 200 snapshots are taken to predict the following 800 snapshots. On the other hand, more snapshots do not necessarily guarantee better prediction results.

However, with the increase in the number of training snapshots, prediction performance did not deteriorate drastically. Hence, it can be inferred that there exists an optimum spot for the choice of data snapshots. We can further infer from figure \ref{fig:8c} and \ref{fig:8d} that irrespective of embedding this trend is followed. In the case of this study, the optimum number of training snapshots was found to be 400. The prediction results verified the correctness of the system identification using the optimum number of training snapshots.\

\subsection{Impact of Prediction Window}
\begin{figure}[H] 
    \centering
  \begin{subfigure}{0.24\textwidth}
    \includegraphics[width=\textwidth]{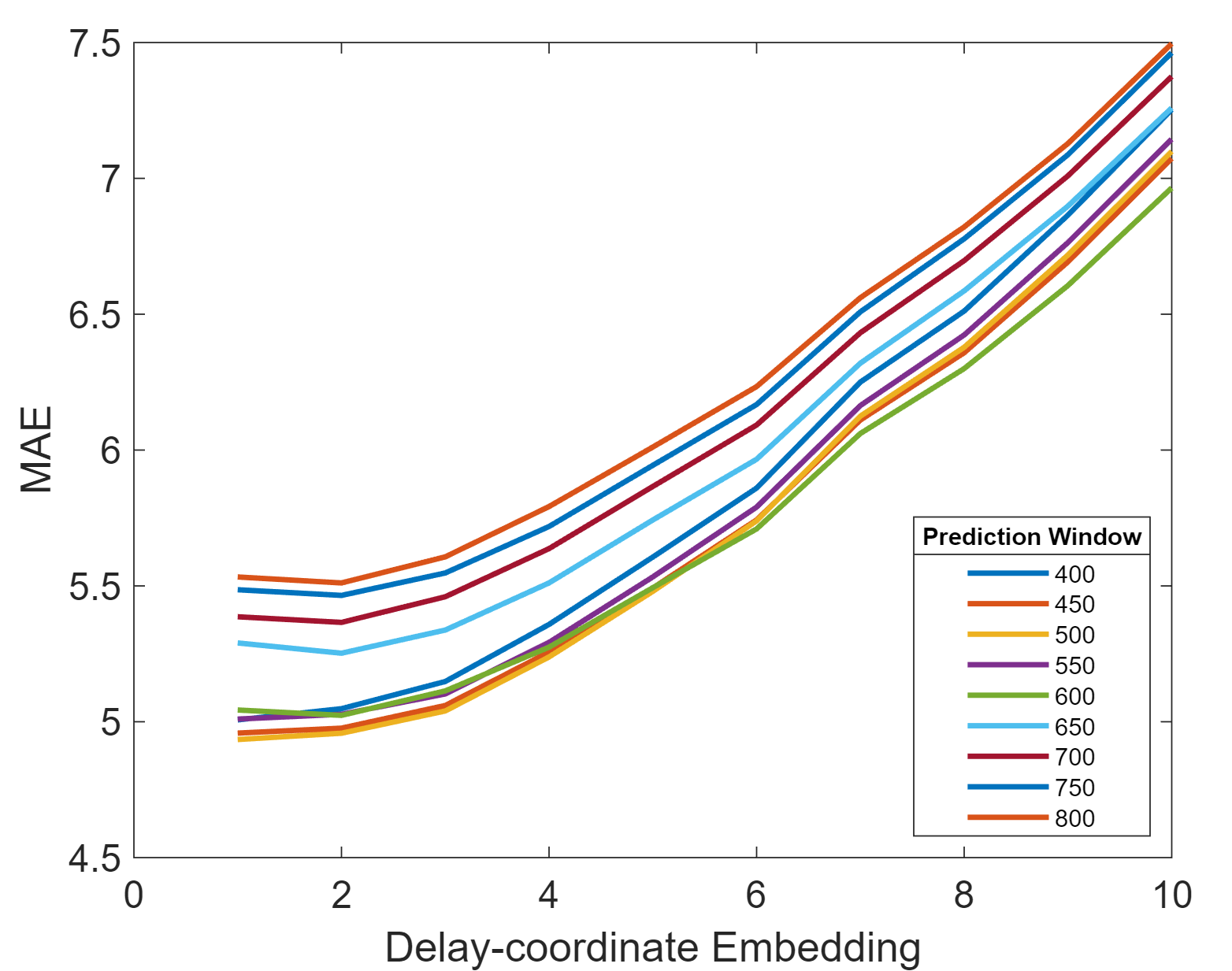}
    \caption{The impact of embedding on prediction results for different prediction windows}
    \label{fig:8a}
  \end{subfigure}%
  ~
  \begin{subfigure}{0.24\textwidth}
    \includegraphics[width=\textwidth]{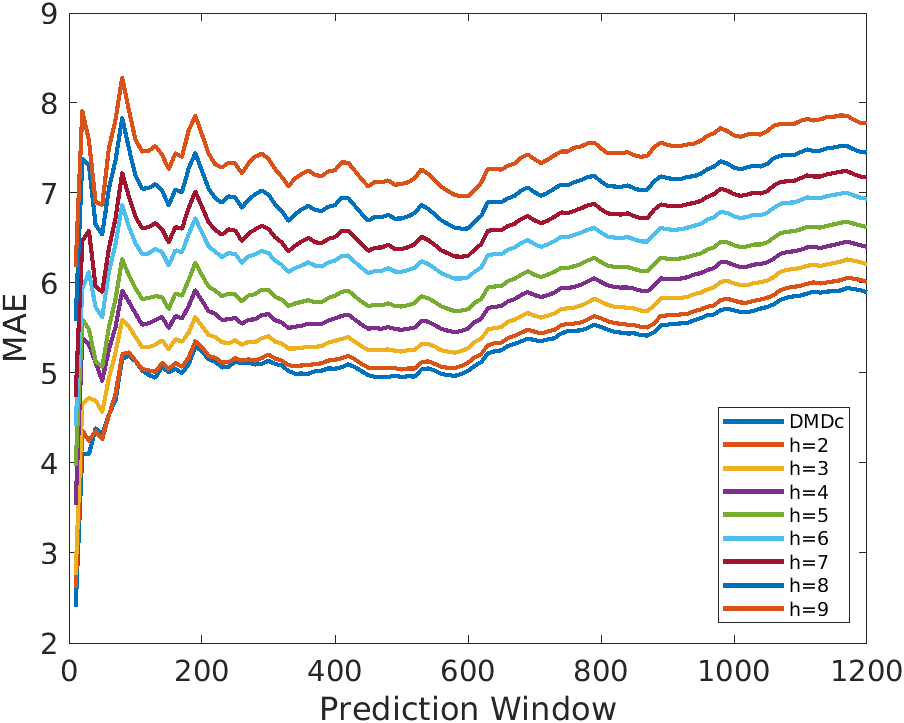}
    \caption{The impact of prediction windows on prediction results for different embedding ($h$)}
    \label{fig:8b}
  \end{subfigure}
   ~
      \begin{subfigure}{0.24\textwidth}
    \includegraphics[width=\textwidth]{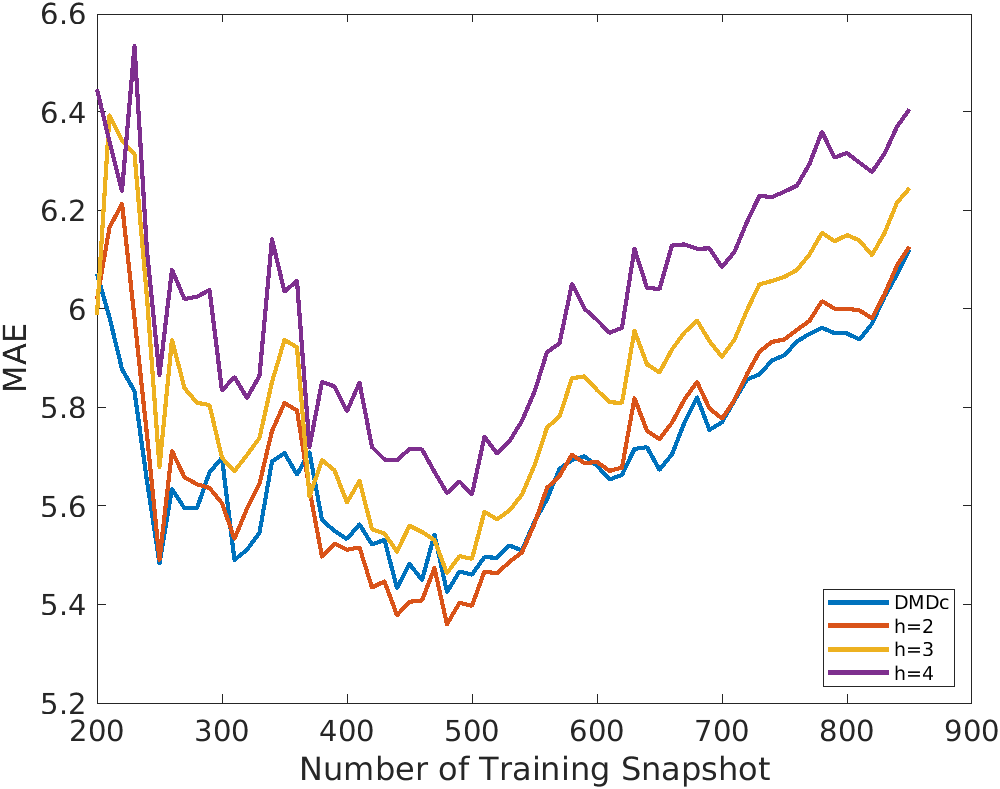}
    \caption{The impact of the choice training snapshot on prediction results for lower number of embedding}
    \label{fig:8c}
  \end{subfigure}%
    ~
    \begin{subfigure}{0.24\textwidth}
    \includegraphics[width=\textwidth]{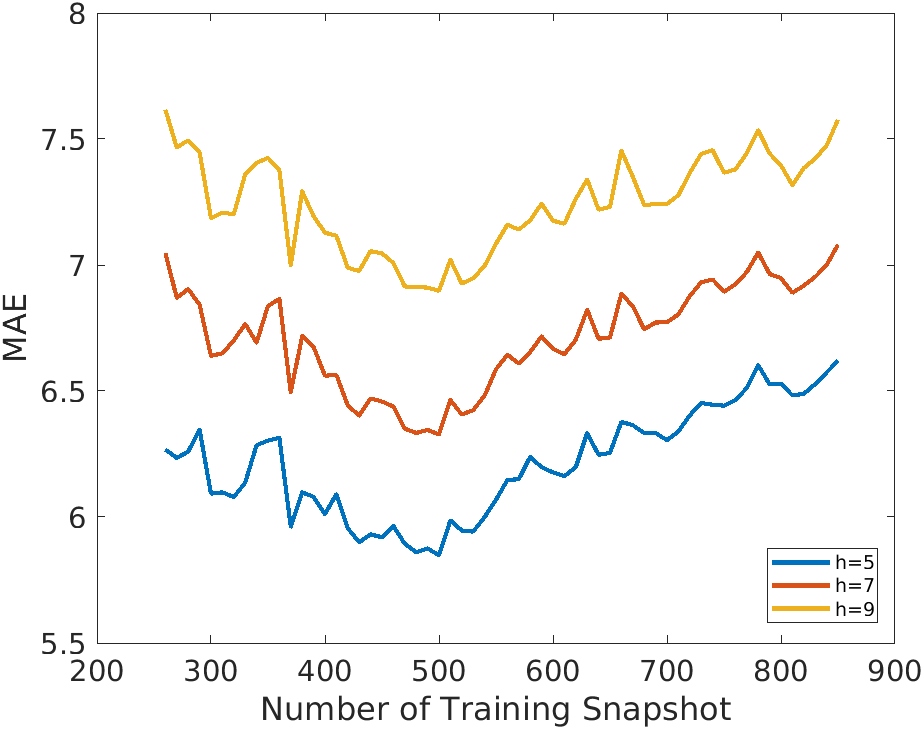}
    \caption{The impact of the choice of training snapshot on prediction for higher number of embedding}
    \label{fig:8d}
    \end{subfigure}
    \caption{The impact of delay-coordinate embedding, prediction window, and the choice of training snapshot on prediction performance at the Rouse intersection}
    \label{fig:8}
    \centering
\end{figure}

\noindent It is expected that for longer prediction windows, the error metrics will increase. Figure \ref{fig:8b} corroborates the assumption. However, the increase in the prediction window does not deteriorate the performance of prediction significantly. Rather the MAE slowly increases over a long range. This trend is followed at different delay-coordinate embedding. 

\section{Validation} \label{sec:performance}
This section aims to evaluate the accuracy of the DMD-based system identification. The evaluation compares the prediction results of DMDc, HDMDc, and LSTM concerning the ground truth, i.e., the actual queue lengths measured at the intersections. Two indexes are suggested in this study, Root mean square error (RMSE), and mean absolute error (MAE). See the equations below:

  \begin{equation}
    RMSE = \sqrt{(\frac{1}{n})\sum_{i=1}^{n}(y_{i} - x_{i})^{2}}\\
  \end{equation}

  \begin{equation}
    MAE = (\frac{1}{n})\sum_{i=1}^{n}\left | y_{i} - x_{i} \right |\\
  \end{equation}

where $y_{i}$ and $x_{i}$ are the actual and predicted queue lengths at the intersections respectively.\\

We played out a few investigations with various batch sizes and hidden layers to check the exactness of the LSTM-NN model for various delays. We found that it performs sensibly well with one hidden layer with 300 neurons and a cluster size of 24. Nonetheless, on the off chance that we increment the number of hidden layers, it takes less time. We tried various blends of hyperparameters lastly fixed them dependent on expectation precision. The last model contains two hidden layers. The primary layer contains 128 neurons, and the subsequent layer contains 64 neurons. To stay away from the overfitting of the model, we added a dropout of 0.1 at the main hidden layer and 0.05 at the second hidden layer. We additionally applied early halting to keep away from overfitting. For spiky information, both Adam and Adagrad analyzer work best, however considering time necessity, the Adam optimizer unites quicker than the Adagrad optimizer. So, we used Adam optimizer. \\

The prediction errors of LSTM, DMDc, and HDMDc are shown in the \ref{table:2}.\\

\begin{table}[!ht]
\caption{Comparing Prediction Error between LSTM, DMDc, and HMDc}
	\begin{center}
\begin{tabular}{|c|c|c|c|c|c|c|c|}
\hline
Intersection                                                                            & Index                                                          & \multicolumn{2}{c|}{LSTM} & \multicolumn{2}{c|}{DMDc} & \multicolumn{2}{c|}{HDMDc} \\ \hline
\multirow{5}{*}{\begin{tabular}[c]{@{}c@{}}Rouse\\ Intersection\end{tabular}}           & \begin{tabular}[c]{@{}c@{}}Training\\ (seconds)\end{tabular}   & \multicolumn{2}{c|}{400}  & \multicolumn{2}{c|}{400}  & \multicolumn{2}{c|}{400}   \\ \cline{2-8} 
                                                                                        & \begin{tabular}[c]{@{}c@{}}Prediction\\ (seconds)\end{tabular} & 200          & 400          & 200          & 400          & 200           & 400          \\ \cline{2-8} 
                                                                                        & RMSE                                                           & 5.20        & 4.50        & 8.07       & 7.73        &  9.18        &     8.73        \\ \cline{2-8} 
                                                                                        & MAE                                                            & 1.27        & 1.06        & 5.21        & 5.00        &     5.80         &   5.60          
                                                                                                  \\ \hline
\multirow{5}{*}{\begin{tabular}[c]{@{}c@{}}Murdock\\ Blvd.\\ Intersection\end{tabular}} & \begin{tabular}[c]{@{}c@{}}Training\\ (seconds)\end{tabular}   & \multicolumn{2}{c|}{400}  & \multicolumn{2}{c|}{400}  & \multicolumn{2}{c|}{400}   \\ \cline{2-8} 
                                                                                        & \begin{tabular}[c]{@{}c@{}}Prediction\\ (seconds)\end{tabular} & 200          & 400          & 200          & 400          & 200           & 400          \\ \cline{2-8} 
                                                                                        & RMSE                                                           & 1.246         & 0.99        & 6.17        & 7.28        &    3.40          &  7.35           \\ \cline{2-8} 
                                                                                        & MAE                                                            & 0.31         & 0.20        & 3.89         & 4.33         &       1.54       &   4.37                   \\ \hline
\end{tabular}
\end{center}
\label{table:2}
\end{table}
It is observed from the Table \ref{table:2} that DMDc, and HDMDc performed closely to LSTM for (200 seconds and 400 seconds prediction window) queue length prediction in the intersections. LSTM is known for accomplishing predictions with high accuracy for time series data. In this comparison, we aimed to evaluate the system identification of the DMDc, and HDMDc algorithms. The results in the table \ref{table:2} show that DMDc and HDMDc were able to predict the queue lengths with an accuracy close to LTSM for queue length prediction. So this indicates, the system identification of the nonlinear dynamics of the signalized traffic intersection using DMDc, and HDMDc performed to a satisfactory level. Finally, it can be inferred from the results that DMD based algorithms can perform effectively for nonlinear system identification and short term future state prediction for signalized traffic intersection.

\section{Discussion and Conclusion}\label{sec:con} 
Traffic congestion is a fundamental problem in urban transportation. Coordinated adaptive traffic signal control can play a vital role in reducing traffic congestion. For implementing an effective adaptive traffic signal controller, it is necessary to understand the nonlinear traffic dynamics. However, it is challenging to understand the nonlinear behavior of traffic at a signalized intersection, even harder for an interconnected network of intersections. Moreover, the design of an adaptive controller using nonlinear dynamics is complicated without any stability guarantees. This paper is a step forward for uncovering the underlying dynamics of complex traffic phenomena using data-driven dynamic mode decomposition algorithms. The presented approach is purely data-driven and appropriate for developing new-age dynamic models incorporating sensor data from emerging technologies.

The study demonstrated the Koopman theoretic modeling approach in the context of adaptive signalized intersections. The queue lengths of all the incoming lanes at the intersection were extracted from simulated SUMO trajectories. We explored DMDc and HDMDc to obtain a locally linear system description for queue length dynamics. Using the obtained linear models, we predicted the short-term and long-term queue lengths, which are essential building blocks for any adaptive control design. This demonstrated the applicability as well as served as an implicit measure to evaluate the system identification results. Please note that there is no direct evaluation mechanism available as the actual system dynamics are not known. The prediction results were further analyzed against multiple factors - delay embedding, the effect of the number of training snapshots, and the length of the prediction window. The prediction results were compared with the state-of-the-art LSTM methods. However, it is essential to mention here that the prediction is not the only unique strength of DMD-based algorithms. This research's objective was not to outperform the prediction power of time series methods such as LSTM or ARIMA but to obtain a reasonable linearized system model that can be utilized to perform adaptive control and even incorporate in DTA models.

There are a few limitations of the study that we would like to highlight. Firstly, the queue lengths were not considered for each lane but aggregated for each moving direction. Secondly, the intersections were adjacent to each other; however, we did not consider the mutual interactions among them (and other nearby intersections). The yellow light was clubbed with the green light, which may not reflect reality. We only considered the morning peak; however, it remains to be seen if the results are consistent during other times of the day or week.

The application of DMD and Koopman theory is still in nascent stages, particularly in the domain of data-driven intelligent transportation systems. The results obtained in this study can serve as an initial step to build ITS applications such as coordinated adaptive signalized intersections involving multiple intersections along a corridor. Future work should consider analytical reasoning behind the observed results related to delay embedding and training snapshots. Researchers may also consider investigating an isolated intersection and then treat multiple intersections as one dynamic system.

\section*{Acknowledgment}
The authors would like to acknowledge the Regional Integrated Transportation Information System (RITIS) and Florida Department of Transportation (FDOT) to provide the detector data and signal retiming data.

\ifCLASSOPTIONcaptionsoff
  \newpage
\fi
\bibliographystyle{IEEEtran}
\bibliography{trb_template.bib}
\vskip 0pt plus -1fil
\begin{IEEEbiography}[{\includegraphics[width=1in,height=1.25in,clip,keepaspectratio]{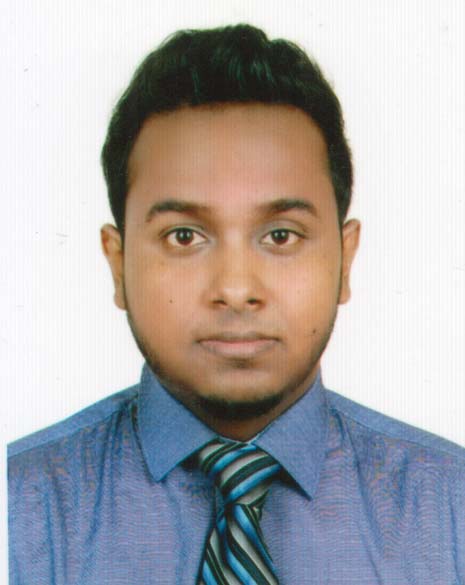}}]%
{Kazi Redwan Shabab} is currently a Ph.D. student at the University of Central Florida. His major is Transportation Engineering. He completed his Bachelor in Urban and Regional Planning from Khulna University of Engineering and Technology in 2018. His research interests include microscopic traffic simulation, traffic safety analysis, connected vehicles and infrastructure, intelligent transportation systems, and computer vision application in Transportation Engineering.
\end{IEEEbiography}
\vskip 0pt plus -1fil
\begin{IEEEbiography}[{\includegraphics[width=1in,height=1.25in,clip,keepaspectratio]{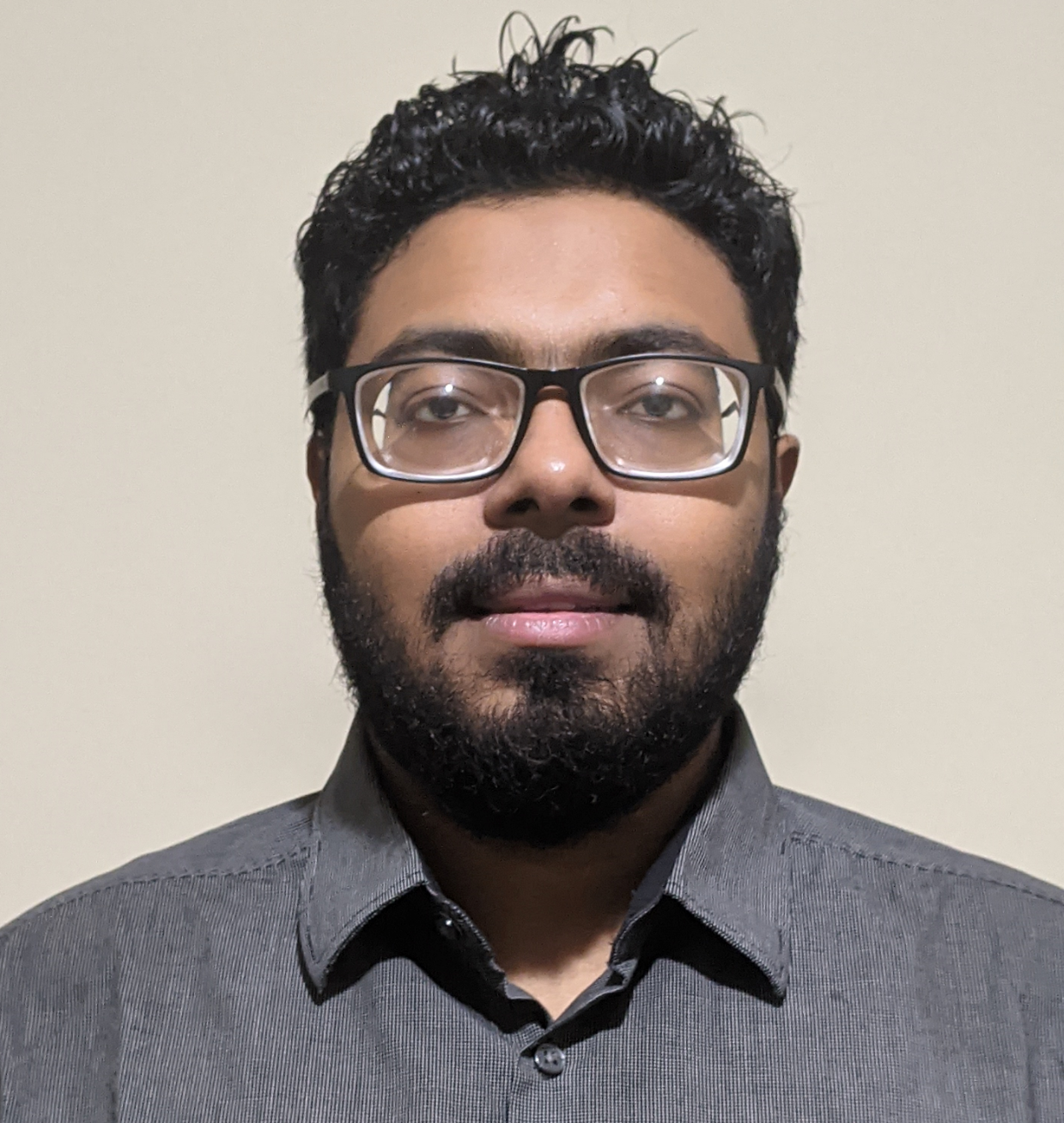}}]%
{Shakib Mustavee}
is a PhD student in the Civil, Environmental, and Construction Engineering Department at the Univerisity  of  Central  Florida. He  is working in the Urban Intelligence and Smart City (URBANITY)Lab under the supervision of Dr. Shaurya Agarwal. He completed his BSc. in Electrical and Electronic Engineering from Bangladesh University of Engineering and Technology in 2017. Currently, his research interests include the application of Dynamic Mode Decomposition and Koopman theory in urban data science.        
\end{IEEEbiography}
\vskip 0pt plus -1fil
\begin{IEEEbiography}[{\includegraphics[width=1in,height=1.25in,clip,keepaspectratio]{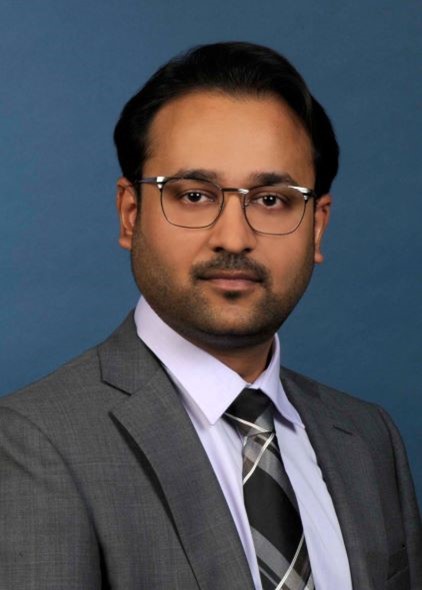}}]%
{Dr. Shaurya Agarwal}
is currently (2018-present) an Assistant Professor in the Civil, Environmental, and Construction Engineering Department at the University of Central Florida. He is the founding director of Urban Intelligence and Smart City (URBANITY) Lab. He was previously (2016-18) an Assistant Professor in Electrical and Computer Engineering Department at the California State University, Los Angeles. He completed his post-doctoral research at New York University (2016) and Ph.D. in Electrical Engineering from University of Nevada, Las Vegas (2015). His B.Tech. degree is in Electronics and Communication Engineering from Indian Institute of Technology (IIT), Guwahati. His research focuses on interdisciplinary areas of cyber-physical systems, smart and connected transportation, and connected and autonomous vehicles. Passionate about cross-disciplinary research, he integrates control theory, information science, data-driven techniques, and mathematical modeling in his work. He has published over 23 peer-reviewed publications and multiple conference papers on various topics including intelligent transportation systems, modeling and control of cyber-physical systems, and information spread on social media. His work has been funded by several private and government agencies.
\end{IEEEbiography}
\vskip 0pt plus -1fil
\begin{IEEEbiography}[{\includegraphics[width=1.1in,height=1.15in,clip,keepaspectratio]{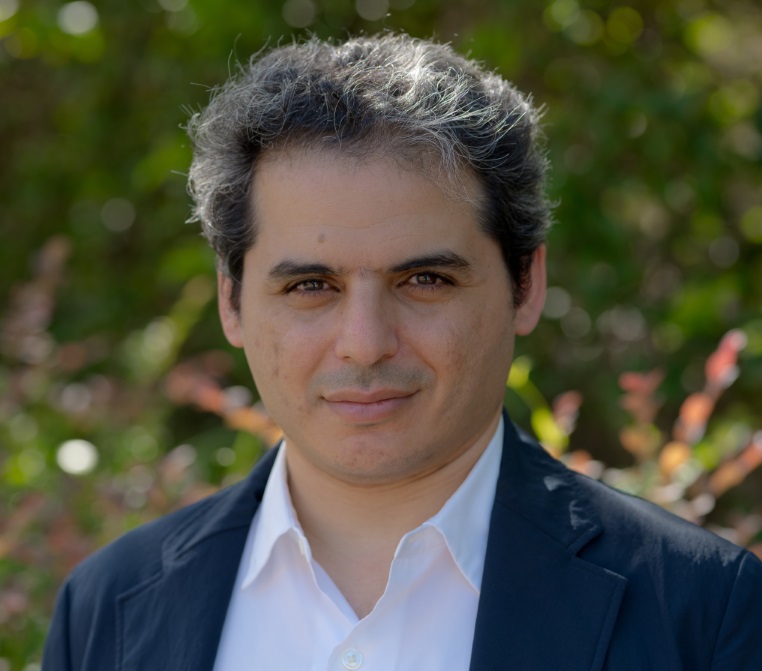}}]
  {Dr. Mohamed H. Zaki}  is an Assistant Professor of Transportation Engineering in the Civil, Environmental \& Construction Engineering Department at the University of Central Florida. Before, Dr. Zaki was a research associate at the Bureau of Intelligent Transportation Systems and Freight Security at the University of British Columbia. He received his Doctoral degrees at the Hardware Verification Group, Concordia University, Montreal, in 2008. Dr. Zaki multidisciplinary research focuses on solving tomorrow's smart cities problems; from the computing and information to its facilities infrastructure. Dr. Zaki studies road safety and road-users' behavior through the automated analysis of traffic data.  Dr. Zaki is an IEEE member and serves on the Transportation Research Board (TRB) ABJ70 Committee on Artificial Intelligence and Advanced Computing Applications.
\end{IEEEbiography}
\vskip 0pt plus -1fil
\begin{IEEEbiography}[{\includegraphics[width=1.1in,height=1.15in,clip,keepaspectratio]{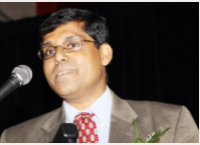}}]
  {Dr. Sajal Das}  is a Professor of Computer Science Department at the Missouri University of Science and Technology. His areas of interest are Cyber-Physical Systems; Security and Privacy; Smart Environments (Smart City, Smart Grid, Smart Healthcare); IoTs; Wireless and Sensor Networks; Mobile and Pervasive Computing; Big Data Analytics; Parallel, Distributed, and Cloud Computing; Social Networks; Systems Biology; Applied Graph Theory and Game Theory.
\end{IEEEbiography}


\end{document}